\documentstyle[epsfig,12pt]{article}
\input{psfig}
\input{epsf}

\def\simge{\mathrel{%
   \rlap{\raise 0.511ex \hbox{$>$}}{\lower 0.511ex \hbox{$\sim$}}}}
\def\simle{\mathrel{
   \rlap{\raise 0.511ex \hbox{$<$}}{\lower 0.511ex \hbox{$\sim$}}}}
 
\def\slashchar#1{\setbox0=\hbox{$#1$}           
   \dimen0=\wd0                                 
   \setbox1=\hbox{/} \dimen1=\wd1               
   \ifdim\dimen0>\dimen1                        
      \rlap{\hbox to \dimen0{\hfil/\hfil}}      
      #1                                        
   \else                                        
      \rlap{\hbox to \dimen1{\hfil$#1$\hfil}}   
      /                                         
   \fi}

\def\thc{\theta_C}
\def\thy{\theta_Y}
\def\dag{\dagger}
\def\nn{\nonumber}
\def\ts{\thinspace}

\def\ra{\rightarrow}

\def\ol{\bar}

\def\be{\begin{equation}} 
\def\ee{\end{equation}} 
\def\bea{\begin{eqnarray}}
\def\eea{\end{eqnarray}}
\def\ba{\begin{array}}
\def\ea{\end{array}}

\def\CD{{\cal D}}

\def\CH{{\cal H}}

\def\CL{{\cal L}}
\def\CM{{\cal M}}

\def\CO{{\cal O}}

\def\CU{{\cal U}}

\def\CZ{{\cal Z}}
\def\chtct{\CH_{TC2}}

\def\chetc{\CH_{ETC}}

\def\atc{\alpha_{TC}}

\def\atro{\alpha_{\tro}}

\def\Ntc{N_{TC}}
\def\suc{SU(3)}

\def\sutc{SU(\Ntc)}

\def\getc{g_{ETC}}

\def\uone{U(1)_1}
\def\utwo{U(1)_2}
\def\uy{U(1)_Y}
\def\suone{SU(3)_1}
\def\sutwo{SU(3)_2}
\def\thw{\theta_W}
\def\kslash{\raise.15ex\hbox{/}\kern-.57em k}
\def\LTC{\Lambda_{TC}}
\def\LETC{\Lambda_{ETC}}
\def\METC{M_{ETC}}
\def\tro{\rho_{T}}

\def\tom{\omega_T}
\def\tpi{\pi_T}

\def\Mv{M_{V_8}}
\def\Mzp{M_{Z'}}
\def\condtbt{\langle \bar t t\rangle}

\def\condtc{{\langle \ol T T \rangle}_{TC}}
\def\condetc{{\langle \ol T T \rangle}_{ETC}}

\def\mev{{\rm MeV}}
\def\gev{{\rm GeV}}
\def\tev{{\rm TeV}}

\def\half{{\textstyle{ { 1\over { 2 } }}}}

\def\fourth{{\textstyle{ { 1\over { 4 } }}}}

\def\nin{\noindent}

\begin{document}
\title{
\vskip -15mm
\begin{flushright}
\vskip -15mm
{\small FERMILAB-PUB-02/040-T\\
BUHEP-02-15\\
hep-ph/0202255\\}
\vskip 5mm
\end{flushright}
{\Large{\bf Two Lectures on Technicolor}}\\ 
}
\author{
\centerline{{Kenneth Lane\thanks{lane@physics.bu.edu}}}\\
\centerline{{Fermi National Accelerator Laboratory}}\\
\centerline{{P.O. Box 500, Batavia, IL 60510}}\\
\centerline{and}\\
\centerline{{Department of Physics, Boston University}}\\
\centerline{{590 Commonwealth Avenue, Boston, MA 02215\footnote{Permanent
address.}}}\\
}
\maketitle
\begin{abstract}
These two lectures on technicolor and extended technicolor (ETC) were
presented at l'Ecole de GIF at LAPP, Annecy-le-Vieux, France, in
September~2001. In Lecture~I, the motivation and structure of this theory of
dynamical breaking of electroweak and flavor symmetries is summarized. The
main phenomenological obstacles to this picture---flavor--changing neutral
currents, precision electroweak measurements, and the large top--quark
mass---are reviewed. Then, their proposed resolutions---walking technicolor
and topcolor--assisted technicolor are discussed. In Lecture~II, a scenario
for CP violation is presented based on vacuum alignment for technifermions
and quarks. It has the novel feature of CP--violating phases that are
rational multiples of $\pi$ to better than one part in $10^{10}$ {\it
without} fine--tuning of parameters. The scheme thereby avoids light axions
and a massless up quark. The mixing of neutral mesons, the mechanism of
top--quark mass generation, and the CP--violating parameters $\epsilon$ and
$\sin(2\beta)$ strongly constrain the form of ETC--generated quark mass
matrices.
\end{abstract}


\newpage

\section*{I. INTRODUCTION TO TECHNICOLOR\footnote{This lecture closely
    follows my lectures at the Frascati 2000 Spring
    School~\cite{frascati}. For other recent reviews, see
    Refs.~\cite{rscreview,cthehs}.}}

\vskip0.1truein

\hskip0.68truein   ``Faith'' is a fine invention

\hskip0.5truein    When Gentlemen can see ---

\hskip0.5truein    But  {\it Microscopes}  are prudent

\hskip0.5truein    In an Emergency.

\vskip0.25truein

\hskip1.0truein --- Emily Dickinson, 1860

\bigskip
\section*{I.1 The Motivation for Technicolor and \hfil\break
Extended Technicolor}

The elements of the standard model of elementary particles have been in place
for more than 25~years now. These include the $\suc\otimes SU(2) \otimes
U(1)$ gauge model of strong and electroweak
interactions~\cite{smtheory,smexpt}. And, they include the Higgs mechanism
used to break {\it spontaneously} electroweak $SU(2)\otimes U(1)$ down to the
$U(1)$ of electromagnetism~\cite{higgs}. In the standard model, couplings of
the elementary Higgs scalar bosons also break {\it explicitly} quark and
lepton chiral--flavor symmetries, giving them hard masses (i.e., mass terms
that appear in the Lagrangian). In this quarter century, the standard model
has stood up to the most stringent experimental tests~\cite{pdg,SM}. The only
clear indications we have of physics beyond this framework are the existence
of neutrino mixing and, presumably, masses (though some would say this
physics is accommodated within the standard model); the enormous range of
masses, about $10^{12}$, between the neutrinos and the top quark; the need
for a new source of CP--violation to account for the baryon asymmetry of the
universe; the likely presence of cold dark matter; and, possibly, a very
small, but nonzero, cosmological constant. These hints are powerful. But they
are also obscure, and they do not point unambiguously to any particular
extension of the standard model.

In addition to these experimental facts, considerable theoretical discomfort
and dissatisfaction with the standard model have dogged it from the
beginning. All of it concerns the elementary Higgs boson picture of
electroweak and flavor symmetry breaking---the cornerstone of the standard
model. In particular:

\begin{enumerate}

\item{} Elementary Higgs models provide no {\it dynamical} explanation for
electroweak symmetry breaking.

\item{} Elementary Higgs models are {\it unnatural}, requiring fine tuning of
parameters to enormous precision.

\item{} Elementary Higgs models with grand unification have a {\it hierarchy}
  problem of widely different energy scales.

\item{} Elementary Higgs models are {\it trivial}.

\item{} Elementary Higgs models provide no insight to {\it flavor} physics.

\end{enumerate}

In nonsupersymmetric Higgs models, there is no explanation why electroweak
symmetry breaking occurs and why it has the energy scale of 1~TeV. The Higgs
doublet self--interaction potential is $V(\phi) = \lambda\ts (\phi^\dagger
\phi - v^2)^2$, where $v$ is the vacuum expectation of the Higgs field $\phi$
{\it provided that} $v^2 \ge 0$. Its experimental value is $v = 2^{-1/4}
G_F^{-1/2} = 246\,\gev$. But what dynamics makes $v^2 > 0$? What dynamics
sets its magnitude?  In supersymmetric Higgs models, the large top--quark
Yukawa coupling can drive $v^2$ positive (by driving $M^2_H$ negative), but
this just replaces one problem with another since we don't know why the top's
Yukawa coupling is $\CO(1)$. Furthermore, this electroweak symmetry breaking
scenario requires the supersymmetric ``mu--term'' to be $\CO(1\,\tev)$. Why
that value?

Elementary Higgs boson models are unnatural. The Higgs boson's mass, $M^2_H =
2 \lambda v^2$ is {\it quadratically} unstable against radiative
corrections~\cite{natural}. Thus, there is no natural reason why $M_H$ and
$v$ should be much less than the energy scale at which the essential physics
of the model changes, e.g., a unification scale or the Planck scale of
$10^{16}\,\tev$. To make $M_H$ very much less that $M_P$, say 1~TeV, the
bare Higgs mass must be balanced against its radiative corrections to the
fantastic precision of a part in $M^2_P/M^2_H \sim 10^{32}$.

In grand--unified Higgs boson models, supersymmetric or not, there are two
very different scales of gauge symmetry breaking, the GUT scale of about
$10^{16}\,\gev$ and the electroweak scale of a few 100~GeV. This hierarchy is
put in by hand, and must be maintained by unnaturally--fine tuning in
ordinary Higgs models, or by the ``set it and forget it'' nonrenormalization
feature of supersymmetry.

Next, taken at face value, elementary Higgs boson models are free field
theories~\cite{trivial}. To a good approximation, the self--coupling
$\lambda(\mu)$ of the minimal one--doublet Higgs boson at an energy scale
$\mu$ is given by
\be\label{eq:lamtriv}
\lambda(\mu) \cong {\lambda(\Lambda) \over {1 + (24 /16 \pi^2)\ts
\lambda(\Lambda) \ts \log (\Lambda /\mu)}} \ts.
\ee
This coupling vanishes for all $\mu$ as the cutoff $\Lambda$ is taken to
infinity, hence the description ``trivial''. This feature persists in a
general class of two--Higgs doublet models~\cite{rscdk} and it is probably
true of all Higgs models. Triviality really means that elementary--Higgs
Lagrangians are meaningful only for scales $\mu$ below some cutoff
$\Lambda_\infty$ at which new physics sets in. The larger the Higgs couplings
are, the lower the scale $\Lambda_\infty$. This relationship translates into
the so--called triviality bounds on Higgs masses. For the minimal model, the
connection between $M_H$ and $\Lambda_\infty$ is
\be\label{eq:Mtriv}
M_H(\Lambda_\infty) \cong \sqrt{2 \lambda(M_H)} \ts v = {2 \pi v \over
{\sqrt{3 \log (\Lambda_\infty/M_H)}}} \ts.
\ee
Clearly, the cutoff has to be greater than the Higgs mass for the effective
theory to have some range of validity. From lattice--based
arguments~\cite{trivial}, $\Lambda_\infty \simge 2 \pi M_H$. Since $v$ is
fixed at 246~GeV in the minimal model, this implies the triviality bound $M_H
\simle 700\,\gev$.\footnote{Precision electroweak measurements suggesting
that $M_H < 200\,\gev$ do not take into account additional interactions that
occur if the Higgs is heavy and the scale $\Lambda$ relatively low. Chivukula
and Evans have argued that these interactions allow $M_H = 400$--$500\,\gev$
to be consistent with the precision measurements~\cite{rscne}.} If the
standard Higgs boson were to be found with a mass this large or larger, we
would know for sure that additional new physics is lurking in the range of a
few~TeV. If the Higgs boson is light, less than 200--300~GeV, as it is
expected to be in supersymmetric models, this transition to a more
fundamental theory may be postponed until very high energy, but what lies up
there worries us nonetheless.

Finally, in all elementary Higgs models, supersymmetric or not, every aspect
of flavor is completely mysterious, from the primordial symmetry defining the
number of quark and lepton generations to the bewildering patterns of flavor
breaking. The presence of Higgs bosons has no connection to the existence of
multiple and identical fermion generations. The flavor--symmetry breaking
Yukawa couplings of Higgs bosons to fermions are arbitrary free parameters,
put in by hand. As far as we know, this is a logically consistent state of
affairs, and we may not understand flavor until we understand the physics of
the Planck scale. I do not believe this. And, I cannot see how this problem,
more pressing and immediate than any save electroweak symmetry breaking
itself, can be so cavalierly set aside by those pursuing the ``theory of
everything''.\footnote{This is not quite fair. In the early days of the
second string revolution, in the mid 1980s, there was a great deal of hope
and even expectation that string theory would provide the spectrum---quantum
numbers and masses---of the quarks and leptons. Those string pioneers and
their descendants have learned how hard the flavor problem is. Indeed, one of
them bet me in 1985 that string theory would produce the quark mass matrix by
1990. That wager cost him lunch for my wife and me at Marc Veyrat's!}

The dynamical approach to electroweak and flavor symmetry breaking known as
technicolor (TC)~\cite{tc,kltasi,rscreview} and extended technicolor
(ETC)~\cite{etceekl,etcsd} emerged in the late 1970s in response to these
shortcomings of the standard model. This picture was motivated first of all
by the premise that {\it every} fundamental energy scale should have a
dynamical origin. Thus, the weak scale embodied in the Higgs vacuum
expectation value $v = 246\,\gev$ should reflect the characteristic energy of
a new strong interaction---technicolor---just as the pion decay constant
$f_\pi = 93\,\mev$ reflects QCD's scale $\Lambda_{QCD} \sim 200\,\mev$. For
this reason, I write $F_\pi = 2^{-1/4} G_F^{-1/2} = 246\,\gev$ to emphasize
that this quantity has a dynamical origin.

Technicolor, a gauge theory of fermions with no elementary scalars, is
modeled on the precedent of QCD: The electroweak assignments of quarks to
left--handed doublets and right--handed singlets prevent their bare mass
terms. If there are no elementary Higgses to couple to, quarks have a large
chiral symmetry, $SU(6)_L \otimes SU(6)_R$ for three generations. This
symmetry is spontaneously broken to the diagonal (vectorial) $SU(6)$ subgroup
when the QCD gauge coupling grows strong near $\Lambda_{QCD}$. This produces
35 massless Goldstone bosons, the ``pions''. According to the Higgs
mechanism---whose operation requires no {\it elementary} scalar
bosons~\cite{jjcn}---this yields weak boson masses of $M_W = M_Z\cos\thw =
\half \sqrt{3} g f_\pi \simeq 50\,\mev$~\cite{tc}. These masses are 1600
times too small, but they do have the right ratio. Suppose, then, that there
are fermions belonging to a complex representation of a new gauge group,
technicolor (taken to be $\sutc$), whose coupling $\atc$ becomes strong at
$\LTC = 100$s of GeV. If, like quarks, technifermions form left--handed
doublets and right--handed singlets under $SU(2) \otimes U(1)$, then they
have no bare masses. When $\atc$ becomes strong, the technifermions' chiral
symmetry is spontaneously broken, Goldstone bosons appear, three of them
become the longitudinal components of $W^\pm$ and $Z^0$, and their masses are
$M_W = M_Z\cos\thw = \half g F_\pi$. Here, $F_\pi \sim \LTC$ is the decay
constant of the linear combination of the absorbed ``technipions''.

Technicolor, like QCD, is asymptotically free. This solves in one stroke the
naturalness, hierarchy, and triviality problems! The mass of all
ground--state technihadrons, including Higgs--like scalars (though that
language is neither accurate nor useful in technicolor) is of order $\LTC$ or
less. There are no large renormalizations of bound state masses, hence no
fine tuning of parameters. If the technicolor gauge symmetry is embedded at
a very high energy $\Lambda$ in some grand unified gauge group with a
relatively weak coupling, then the characteristic energy scale $\LTC$---where
the coupling $\atc$ becomes strong enough to trigger chiral symmetry
breaking---is naturally exponentially smaller than $\Lambda$. Finally,
asymptotically free field theories are nontrivial. A minus sign in the
denominator of the analog of Eq.~(\ref{eq:lamtriv}) for $\atc(\mu)$ prevents
one from concluding that it tends to zero for all $\mu$ as the cutoff is
taken to infinity. No other scenario for the physics of the TeV scale solves
these problems so neatly. Period.

Technicolor alone does not address the flavor problem. It does not tell us
why there are multiple generations and it does not provide explicit breaking
of quark and lepton chiral symmetries. Something must play the role of Higgs
bosons to communicate electroweak symmetry breaking to quarks and leptons.
Furthermore, in all but the minimal TC model with just one doublet of
technifermions, there are Goldstone bosons, technipions $\tpi$, in addition
to $W_L^\pm$ and $Z_L^0$. These must be given mass and their masses must be
more than 50--100~GeV for them to have escaped detection. Extended
technicolor (ETC) was invented to address {\it all} these aspects of flavor
physics~\cite{etceekl}. It was also motivated by the desire to make flavor
understandable at energies well below the GUT scale solely in terms of {\it
gauge} dynamics of the kind that worked so neatly for electroweak symmetry
breaking, namely, technicolor.  Let me repeat: the ETC approach is based on
the gauge dynamics of fermions only. There can be {\it no elementary scalar
fields} to lead us into the difficulties technicolor itself was invented to
escape.

\section*{I.2 Dynamical Basics}

In extended technicolor, ordinary $SU(3)$ color, $\sutc$ technicolor, and
flavor symmetries are unified into the ETC gauge group, $G_{ETC}$. We then
understand flavor, color, and technicolor as subsets of the quantum numbers
of extended technicolor. Technicolor and color are exact gauge
symmetries. Flavor gauge symmetries are broken at one or more high energy
scales $\LETC \simeq \METC/g_{ETC}$ where $\METC$ is a typical flavor gauge
boson mass.

In these lectures, I assume that $G_{ETC}$ commutes with electroweak
$SU(2)$. In this case, it must {\it not} commute with electroweak $U(1)$,
i.e., some part of that $U(1)$ must be contained in $G_{ETC}$. Otherwise,
there will be very light pseudoGoldstone bosons which behave like classical
axions and are ruled out experimentally~\cite{etceekl,CPreview}. More
generally, all fermions---technifermions, quarks, and leptons---must form no
more than four {\it irreducible} ETC representations: two equivalent ones for
left--handed up and down--type fermions and two inequivalent ones for
right--handed up and down fermions (so that up and down mass matrices are not
identical). In other words, ETC interactions explicitly break all {\it
global} flavor symmetries so that there are no very light pseudoGoldstone
bosons or fermions.\footnote{I leave neutrinos out of this discussion. Their
very light masses are not yet understood in the ETC framework.}

The energy scale of ETC gauge symmetry breaking into $SU(3) \otimes \sutc$ is
high, well above the TC scale of 0.1--1.0~TeV. The broken gauge interactions,
mediated by massive ETC boson exchange, give mass to quarks and leptons by
connecting them to technifermions (Figure~1a). They give mass to technipions by
connecting technifermions to each other (Figure~1b).

The graphs in Figure~1 are convergent: The changes in chirality imply
insertions on the technifermion lines of the momentum--dependent dynamical
mass, $\Sigma(p)$. This function falls off as $1/p^2 \ts (\log(p/\LTC))^c$ in
an asymptotically free theory at weak coupling and, in any case, at least as
fast as $1/p$~\cite{kdlhdp,agchmg}. For such a power law, the dominant
momentum running around the loop is $\CO(\METC)$ (extending down to
$\Lambda_{TC}$ for a $1/p^2$ falloff of $\Sigma$). Then, the operator product
expansion tells us that the generic quark or lepton mass and technipion mass
are given by the expressions
\begin{figure}[t]
 \vspace{6.0cm}
\includegraphics{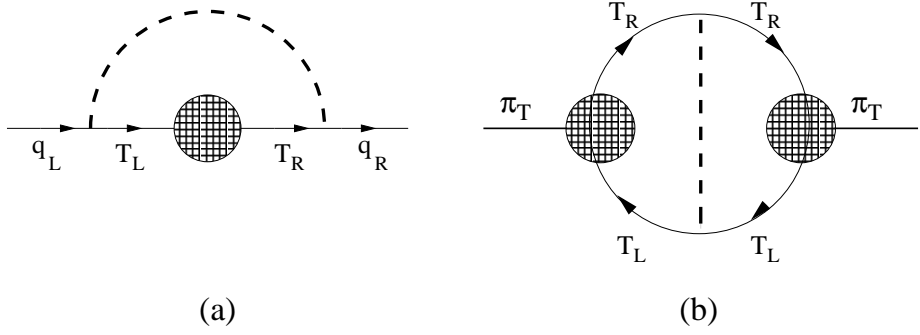}
\vskip-1.5truecm
 \caption{\it
   Graphs for ETC generation of masses for (a) quarks and leptons and (b)
   technipions. The dashed line is a massive ETC gauge boson. Higher--order
   technicolor gluon exchanges are not indicated; from Ref.~\cite{etceekl}.
    \label{fig1} }
\end{figure}
\bea\label{eq:masses}
& & m_q(\METC) \simeq m_\ell(\METC)  \simeq 2{g_{ETC}^2 \over
{M_{ETC}^2}} \langle \ol T_L T_R \rangle_{ETC} \ts; \\
& & F^2_T M^2_{\pi_T}  \simeq 2\ts {g^2_{ETC} \over {M^2_{ETC}}}
\ts \langle \ol T_L T_R \ol T_R T_L \rangle_{ETC} \ts.
\eea

Here, $m_q(\METC)$ is the quark mass renormalized at $\METC$. It is a hard
mass in that it scales like one (i.e., logarithmically) for energies below
$\METC$. Above that, it falls off more rapidly, like $\Sigma(p)$. The
technipion decay constant $F_T = F_\pi/\sqrt{N}$ in TC models containing $N$
identical electroweak doublets of color--singlet technifermions. The vacuum
expectation values $\langle \ol T_L T_R \rangle_{ETC}$ and $\langle \ol T_L
T_R \ol T_R T_L \rangle_{ETC}$ are the bilinear and quadrilinear
technifermion condensates renormalized at $\METC$. The bilinear condensate is
related to the one renormalized at $\LTC$, expected by scaling from QCD to be
\be\label{eq:ctc}
\langle \ol T_L T_R \rangle_{TC} = \half \condtc \simeq 2 \pi F^3_T \ts,
\ee
by the equation
\be\label{eq:condrenorm}
\condetc = \condtc \ts \exp\left(\int_{\LTC}^{M_{ETC}} \ts {d \mu
\over {\mu}} \ts \gamma_m(\mu) \right) \ts.
\ee
The anomalous dimension $\gamma_m$ of the operator $\ol T T$ is given in
perturbation theory by
\be\label{eq:gmm}
\gamma_m(\mu) = {3 C_2(R) \over {2 \pi}} \atc(\mu) + O(\atc^2) \ts,
\ee
where $C_2(R)$ is the quadratic Casimir of the technifermion
$\sutc$--representation $R$. For the fundamental representation of $\sutc$,
it is given by $C_2(\Ntc) = (\Ntc^2-1)/2\Ntc$. Finally, in the large--$\Ntc$
approximation (which will be questionable in the walking technicolor theories
we discuss later, but which we adopt anyway for rough estimates)
\be\label{eq:large_n}
\langle \ol T_L T_R \ol T_R T_L \rangle_{ETC} \simeq \langle \ol T_L
T_R\rangle_{ETC} \ts  \langle \ol T_R T_L \rangle_{ETC} = \fourth \condetc^2
\ts.
\ee

We can estimate $\METC$ and $M_{\tpi}$ if we assume that technicolor is
QCD--like~\cite{etceekl}. In that case, its asymptotic freedom sets in
quickly (or ``precociously'') at energies above $\LTC$ and, so,
$\gamma_m(\mu) \ll 1$ for $\mu$ greater than a few times $\LTC$. Then
Eq.~(\ref{eq:ctc}) applies to $\condetc$ as well. For $N$ technidoublets, the
ETC scale required to generate $m_q(\METC) \simeq 1\,\gev$ is
\be\label{eq:letc}
\Lambda_{ETC} \equiv {M_{ETC} \over {g_{ETC}}} \simeq \sqrt{{4 \pi F_\pi^3
\over {m_q N^{3/2}}}} \simeq {14\,\tev\over{N^{3/4}}} \ts.
\ee
This is pretty low, but the estimate is rough. The typical technipion mass
implied by this ETC scale is
\be\label{eq:mpi}
M_{\tpi} \simeq {\condtc \over{\sqrt{2}\Lambda_{ETC} F_T}} \simeq
{40\,\gev\over{N^{1/4}}} \ts.
\ee

Finally, some phenomenological basics: In any model of technicolor, one
expects bound technihadrons with a spectrum of mesons paralleling what we see
in QCD. The principal targets of collider experiments are the spin--zero
technipions and spin--one isovector technirhos and isoscalar techniomegas. In
the minimal one--technidoublet model ($T = (T_U,T_D)$), the three technipions
are the longitudinal components $W_L$ of the massive weak gauge bosons.
Susskind pointed out that the analog of the QCD decay $\rho \ra \pi\pi$ is
$\tro \ra W_L W_L$~\cite{tc}. In the limit that $M_{\tro} \gg M_{W,Z}$, the
equivalence theorem states that the amplitude for $\tro \ra W_L W_L$ has the
same form as the one for $\rho \ra \pi\pi$. If we scale technicolor from QCD
and use large--$\Ntc$ arguments, the strength of this amplitude is $g_{\tro}
\equiv \sqrt{4\pi \atro} \simeq \sqrt{4\pi (3/\Ntc) \alpha_\rho}$ where
$\alpha_\rho = 2.91$. The $\tro$ mass and decay rate are~\cite{ehlq}:
\bea\label{eq:minrhot}
&&M_{\tro} = \sqrt{{3\over{\Ntc}}}\ts {F_\pi\over{f_\pi}} M_\rho
\simeq 2 \sqrt{{3\over{\Ntc}}}\,\tev \ts, \nn\\
&&\Gamma(\tro \ra W_L W_L ) = {2\atro p_W^3\over{3 M^2_{\tro}}} \simeq 500
\left({3\over{N_{TC}}} \right)^{3/2}\,\gev  \ts.
\eea

In the minimal model, a very high energy machine, such as the ill--fated
Superconducting Super Collider or the $200\,\tev$ Very Large Hadron Collider
or a $2\,\tev$ Linear Collider is needed to discover the lightest
technihadrons.\footnote{It is possible that, like the attention paid to
discovering the minimal standard model Higgs boson, this emphasis on the $W_L
W_L$ decay mode of the $\tro$ is somewhat misguided~\cite{kltasi}.  Since the
minimal $\tro$ is so much heavier than $2M_W$, this mode may be suppressed by
the high $W$--momentum in its decay form factor. Then, the $\tro$ decays to
four or more weak bosons may be competitive or even dominant. This means that
the minimal $\tro$ may be wider than indicated in Eq.~(\ref{eq:minrhot}) and,
in any case, that its decays are much more complicated than previously
thought. Furthermore, walking technicolor~\cite{wtc}, discussed below,
implies that the spectrum of technihadrons cannot be exactly
QCD--like. Rather, there must be something like a tower of technirhos
extending almost up to $\METC \simge$ several 100~TeV. Whether or not these
would appear as discernible resonances is an open question~\cite{hemc}. All
these remarks apply as well to the isoscalar $\tom$ and its excitations.} In
nonminimal models, where $N \ge 2$, the signatures of technicolor ought to be
accessible at the Large Hadron Collider (LHC) and at a comparable lepton
collider. As discussed in Ref.~\cite{frascati}, the many technifermions
needed for walking technicolor and topcolor--assisted technicolor (see below)
mean that signatures are likely to be within reach of the Tevatron Collider
in Run~II!~\footnote{I did not have time in these lectures to discuss the
very interesting phenomenology of this ``low--scale technicolor''. I urge
students to consult this reference. This will be an active research program
at the Tevatron Collider over the next five years.}

\section*{I.3 Dynamical Perils}

Technicolor and extended technicolor are challenged by a number of
phenomenological hurdles, but the most widely cited causes of the ``death of
technicolor'' (prematurely announced, like Mark Twain's) are flavor--changing
neutral current interactions (FCNC)~\cite{etceekl,ellisfcnc}, precision
measurements of electroweak quantities (STU)~\cite{pettests}, and the large
mass of the top quark. We discuss these in turn.\footnote{Much of the
discussion here on FCNC and STU is a slightly updated version of that
appearing in my 1993 TASI lectures~\cite{kltasi}.}

\subsection*{I.3.1 \it {Flavor--Changing Neutral Currents}}

Extended technicolor interactions are expected to have flavor--changing
neutral currents involving quarks and leptons. The reason is simple:
Realistic quark mass matrices require ETC transitions between different
flavors, $q \ra T \ra q'$, or technifermion mixing so that $q \ra T \ra T'
\ra q'$. In the first case, there must be ETC currents of the form $\ol
q'_{L,R} \ts \gamma_\mu \ts T_{L,R}$ and $\ol T_{L,R} \ts \gamma_\mu \ts
q_{L,R}$. Their commutator algebra includes the ETC currents $\ol q'_{L,R}
\ts \gamma_\mu \ts q_{L,R}$. In the second case, these currents may be
generation conserving. Either way, ETC interactions necessarily produce $\ol
q q \ol q q$ operators involving light quarks. Similarly, there will be $\ol
q q \ol \ell \ell$ and $\ol \ell \ell \ol \ell \ell$ operators. Even if these
interactions are electroweak--eigenstate conserving (or generation
conserving), they will induce FCNC four--fermion operators after
diagonalization of mass matrices and transformation to the mass--eigenstate
basis. No satisfactory GIM mechanism has ever been found that eliminates
these FCNC interactions~\cite{etcgim}.

The most stringent constraint on ETC comes from $\vert \Delta S \vert = 2$
interactions. Such an interaction has the generic form
\be\label{eq:dstwo}
\CH'_{\vert \Delta S \vert = 2} = {g^2_{ETC} \ts V^2_{ds} \over
{M^2_{ETC}}} \ts\ts \ol d \ts \Gamma^\mu s \ts\ts \ol d \ts \Gamma'_\mu s +
{\rm h.c.}
\ee
Here, $V_{ds}$ is a mixing--angle factor; it may be complex and seems
unlikely to be much smaller in magnitude than the Cabibbo angle, say $0.1
\simle \vert V_{ds} \vert \simle 1$. The matrices $\Gamma_\mu$ and
$\Gamma'_{\mu}$ are left-- and/or right--chirality Dirac matrices. I shall
put $\Gamma_\mu, \ts \Gamma'_\mu = \half \gamma_\mu \ts (1 - \gamma_5)$ and
count the interaction twice to allow for different chirality terms in
$\CH'_{\vert \Delta S \vert = 2}$. The contribution of this interaction to
the $K_L - K_S$ mass difference is then estimated to be
\bea\label{eq:klks}
(\Delta M_K)_{ETC} &\equiv& 2{\rm Re}(M_{12})_{ETC} \nn \\
&=& {4g^2_{ETC} \ts {\rm Re}(V^2_{ds}) \over 8M_{K} {M^2_{ETC}}} \ts \langle
K^0 \vert \ol d \ts \gamma^\mu (1 - \gamma_5) s \ts \ol d \ts \gamma_\mu (1 -
\gamma_5) s \vert \ol K^0 \rangle \nn\\
&\simeq& {g^2_{ETC} \ts {\rm Re}(V^2_{ds}) \over {M^2_{ETC}}} \ts
f^2_K M_K \ts,
\eea
where I used the vacuum insertion approximation $\langle \Omega \vert \ol d
\gamma_\mu \gamma_5 s\vert \ol K^0(p) \rangle = i \sqrt{2} f_K p_\mu$ with
$f_K \simeq 110\,\mev$. This ETC contribution must be less than the measured
mass difference, $\Delta M_K = 3.5 \times 10^{-18}\,\tev$. This gives the
limit
\be\label{eq:dmlimit}
{M_{ETC} \over {g_{ETC} \ts \sqrt{{\rm Re}(V^2_{ds})}}} \simge 1300\,\tev
\ts.
\ee
If $V_{ds}$ is complex, $\CH'_{\vert \Delta S \vert = 2}$ contributes to the
imaginary part of the $K^0 - \ol K^0$ mass matrix. Using ${\rm Im}(M_{12}) =
\sqrt{2} \Delta M_K |\epsilon| \simeq 1.15\times 10^{-20}\,\tev$, the limit
is
\be\label{eq:epslimit}
{M_{ETC} \over {g_{ETC} \ts \sqrt{{\rm Im}(V^2_{ds})}}} \simge 16000\,\tev
\ts.
\ee

If we use these large ETC masses and scale the technifermion condensates in
Eqs.~(3,4) from QCD---i.e., assume the anomalous dimension $\gamma_m$ is
small so that $\langle \ol T T \ol T T\rangle_{ETC} \simeq \condetc^2 \simeq
\condtc^2 \simeq (4 \pi F^3_T)^2$---we obtain quark and lepton and technipion
masses that are 10--1000 times too small, depending on the size of
$V_{ds}$. This is the FCNC problem. It is remedied by the non--QCD--like
dynamics of technicolor with a slowly running gauge coupling, called walking
technicolor. This will be described in the next section.

\subsection*{I.3.2 \it {Precision Electroweak Measurements}}

Precision electroweak measurements actually challenge technicolor, not
extended technicolor. The basic parameters of the standard $SU(2)\otimes
U(1)$ model---$\alpha(M_Z)$, $M_Z$, $\sin^2 \theta_W$---are measured so
precisely that they may be used to limit new physics at energy scales above
100~GeV~\cite{pettests}. The quantities most sensitive to new physics are
defined in terms of correlation functions of the electroweak currents:
\be\label{eq:pifcn}
\int d^4x \ts e^{-i q\cdot x} \langle\Omega | T\left(j^\mu_i(x)
j^\nu_j(0)\right) | \Omega \rangle =
i g^{\mu\nu} \Pi_{ij}(q^2) + q^\mu q^\nu \ts {\rm terms} \ts.
\ee
Once one has accounted for the contributions from standard model physics,
including a single Higgs boson (whose mass $M_H$ must be assumed), new
high--mass physics affects the $\Pi_{ij}$ functions. Assuming that the scale
of this physics is well above $M_{W,Z}$, it enters the ``oblique'' correction
factors $S$, $T$, $U$ defined by
\bea\label{eq:stu}
&&S= 16\pi {d \over {d q^2}} \left[ \Pi_{33} (q^2) - \Pi_{3Q}(q^2)
\right]_{q^2=0} \ts \equiv \ts 16\pi \left[ \Pi_{33}^{'}(0) - \Pi_{3Q}^{'}(0)
\right] \ts , \nn\\
&&T= {4\pi \over{M^2_Z \cos^2\theta_W \sin^2\theta_W}}
\ts \left[ \Pi_{11}(0) - \Pi_{33}(0) \right] \ts , \nn\\
&&U= 16\pi \left[ \Pi_{11}^{'}(0) - \Pi_{33}^{'}(0) 
\right] \ts .
\eea
The parameter $S$ is a measure of the splitting between $M_W$ and $M_Z$
induced by weak--isospin conserving effects; the $\rho$--parameter is given
by $\rho \equiv M^2_W/M^2_Z \cos^2\thw = 1 + \alpha T$; the $U$--parameter
measures weak--isospin breaking in the $W$ and $Z$ mass splitting. The
experimental limits on $S,T,U$ are~\cite{pdg}
\bea\label{eq:stuvalues}
&&S= -0.07 \pm 0.11 \ts\ts (-0.09)\ts,\nn\\
&&T= -0.10 \pm 0.14 \ts\ts (+0.09)\ts,\nn\\
&&U= +0.11\pm 0.15  \ts\ts (+0.01)\ts.
\eea
The central values assume $M_H = 100\,\gev$, and the parentheses contain the
change for $M_H = 300\,\gev$. The $S$ and $T$--parameters and $M_H$ cannot be
obtained simultaneously from data because the Higgs loops behave
approximately like oblique effects.

The $S$--parameter is the one most touted as a show--stopper for technicolor
\cite{pettests,tctests}. A value of $\CO(1)$ is obtained in technicolor by
scaling up from QCD. For example, for $N$ color--singlet technidoublets,
Peskin and Takeuchi found the positive result
\be\label{eq:svalue}
S = 4 \pi \left(1 + {M^2_{\rho_T} \over{M^2_{a_{1T}}}}\right ) {F^2_\pi \over
{M^2_{\rho_T}}} \simeq 0.25 \ts N \left({N_{TC}\over{3}}\right) \ts.
\ee
The resolution to this problem may also be found in walking technicolor. One
thing is sure: naive scaling of $S$ from QCD is unjustified and probably
incorrect in walking gauge theories. No reliable estimate exists because no
data on walking gauge theories are available to put into the calculation of
$S$.

\subsection*{I.3.3 \it {The Top Quark Mass}}

The ETC scale required to produce $m_t = 175\,\gev$ in Eq.~(3) is
$1.0\,\tev/N^{3/4}$ for $N$ technidoublets. This is uncomfortably close to
the TC scale itself. In effect, TC gets strong and ETC broken at the same
energy; the representation of broken ETC interactions as contact operators is
wrong; and all our mass estimates are questionable. It is possible to raise
the ETC scale so that it is considerably greater than $m_t$, but then one
runs into the problem of fine--tuning the ETC coupling $g_{ETC}$ (just as in
the Nambu--Jona-Lasinio (NJL) model, where requiring the dynamical fermion
mass to be much less than the four--fermion mass scale $\Lambda$ requires
fine--tuning the NJL coupling very close to $4\pi$)~\cite{setc}. This flouts
our cherished principle of naturalness, and we reject it. Another, more
direct, problem with ETC generation of the top mass is that there must be
large weak isospin violation to raise it so high above the bottom mass. This
adversely affects the $\rho$ parameter~\cite{tombowick}. The large effective
ETC coupling to top quarks also makes a large, unwanted contribution to the
$Z \ra \ol b b$ decay rate, in conflict with experiment~\cite{zbbth}.

In the end, there is no plausible way to understand the top quark's large
mass from ETC. Something more is needed. The best idea so far is
topcolor--assisted technicolor~\cite{tctwohill}, in which a new gauge
interaction, topcolor~\cite{topcref}, becomes strong near 1~TeV and generates
a large $\ol t t$ condensate and top mass. This, too, will be described in
the next section.

\section*{I.4 Dynamical Rescues}

The FCNC and STU difficulties of technicolor have a common basis: the
assumption that technicolor is a just a scaled--up version of QCD. Let us
focus on Eqs.(3,4,6), the key equations of extended technicolor. In a
QCD--like technicolor theory, asymptotic freedom sets in quickly above
$\LTC$, the anomalous dimension $\gamma_m \ll 1$, and $\condetc \simeq
\condtc$. The conclusion that fermion and technipion masses are one or more
orders of magnitude too small then followed from the FCNC requirement in
Eqs.~(\ref{eq:dmlimit},\ref{eq:epslimit}) that $\METC/g_{ETC} |V_{ds}| \simge
1300\,\tev$. Scaling from QCD also means that the technihadron spectrum is
just a magnified image of the QCD--hadron spectrum, hence that $S$ is too
large for all technicolor models except, possibly, the minimal one--doublet
model with $N_{TC} \simle 4$.

The solution to these difficulties lies in technicolor gauge dynamics that are
distinctly not QCD--like. The only plausible example is one in which
the gauge coupling $\atc(\mu)$ evolves slowly, or ``walks'', over the large
range of energy $\LTC \simle \mu \simle \METC$~\cite{wtc}. In the extreme
walking limit in which $\atc(\mu)$ is constant, it is possible to obtain an
approximate nonperturbative formula for the $\ol T T$ anomalous dimension
$\gamma_m$, namely,
\be\label{eq:nonpert}
\gamma_m(\mu) = 1 - \sqrt{1-\atc(\mu)/\alpha^*_{TC}} \quad {\rm where}
\ts\ts\ts  \alpha^*_{TC}  = {\pi \over{3 C_2(R)}} \ts.
\ee
This reduces to the expression in Eq.~(\ref{eq:gmm}) for small $\atc$.  It
has been argued that $\gamma_m = 1$ is the signal for spontaneous chiral
symmetry breaking~\cite{agchmg}, and, so, $\alpha^*_{TC}$ is called the
critical coupling for $\chi$SB, with $\pi/3C_2(R)$ its approximate
value.\footnote{An attempt to improve upon this approximation and study its
accuracy is in Ref.~\cite{alm}.} If we define $\LTC$ to be the scale at which
technifermions in the $\sutc$ fundamental representation condense, then
$\atc(\LTC) = \alpha^*_{TC}$.

In walking technicolor, $\atc(\mu)$ is presumed to remain close to its
critical value from $\LTC$ almost up to $\METC$. This implies $\gamma_m(\mu)
\simeq 1$, and by Eq.~(\ref{eq:condrenorm}), the condensate $\condetc$ is
enhanced by a factor of 100 or more. This yields quark masses up to a few GeV
and reasonably large technipion masses despite the very large ETC mass
scale. We will give a numerical example of this in Lecture~II. These enhanced
masses are not enough to account for the top quark; more on that soon.

Another consequence of the walking $\atc$ is that the spectrum of
technihadrons, especially the $I=0,1$ vector and axial vector mesons, $\tro$,
$\tom$, $a_{1T}$ and $f_{1T}$, cannot be
QCD--like~\cite{kltasi,klglasgow,edrta}. In QCD, the lowest lying isovector
$\rho$ and $a_1$ saturate the spectral functions appearing in Weinberg's sum
rules~\cite{sfsr}. Then, the relevant combination $\rho_V - \rho_A$ of
spectral functions falls off like $1/p^6$ for $p > M_{\rho,a_1} \sim
\Lambda_{QCD}$, and the spectral integrals converge very rapidly. This
``vector meson dominance'' of the spectral integrals is related to the
precocious onset of asymptotic freedom in QCD. The $1/p^6$ momentum
dependence is just what one would deduce from a naive, lowest--order
calculation of $\rho_V - \rho_A$ using the asymptotic $1/p^2$ behavior of the
quark dynamical mass $\Sigma(p)$. In walking technicolor, the technifermion's
$\Sigma(p)$ falls only like $1/p^{(2-\gamma_m)} \sim 1/p$ for $\LTC \simle
\METC$, so that $\rho_V - \rho_A \sim 1/p^4$ up to very high energies. To
account for this in terms of spin--one technihadrons, there must be something
like a tower of $\tro$ and $\tom$ extending up to $\METC$. Their mass
spectrum, widths, and couplings to currents cannot be predicted. Without
experimental knowledge of these states, it is impossible to estimate $S$
reliably, any more than it would have been in QCD before the $\rho$ and $a_1$
were discovered and measured.

Finally, another issue that may affect $S$ is that it is usually defined
assuming that the new physics appears at energies well above
$M_{W,Z}$. However, walking technicolor suggests that there are $\tpi $ and
$\tro$ starting near or not far above
$100\,\gev$~\cite{multiklee,elw,tcsm_singlet,frascati}.

We have seen that extended technicolor cannot explain the top quark's large
mass. An alternative approach was developed in the early 90s based on a new
interaction of the third generation quarks.  This interaction, called
topcolor, was invented as a minimal dynamical scheme to reproduce the
simplicity of the one--doublet Higgs model {\it and} explain a very large
top--quark mass~\cite{topcref}. In topcolor, a large top--quark condensate,
$\condtbt$, is formed by strong interactions at the energy scale,
$\Lambda_t$~\cite{topcondref}. To preserve electroweak $SU(2)$, topcolor must
treat $t_L$ and $b_L$ the same. To prevent a large $b$--condensate and mass,
it must violate weak isospin and treat $t_R$ and $b_R$ differently. In order
that the resulting low--energy theory simulate the standard model,
particularly its small violation of weak isospin, the topcolor scale must be
very high---$\Lambda_t \sim 10^{15}\,\gev \gg m_t$. Therefore, this original
topcolor scenario is highly unnatural, requiring a fine--tuning of couplings
of order one part in $\Lambda_t^2/m_t^2 \simeq 10^{25}$ (remember the
finely--tuned Nambu--Jona-Lasinio interaction!).

Technicolor is still the most natural mechanism for electroweak symmetry
breaking, while topcolor dynamics most aptly explains the top mass. Hill
proposed to combine the two into what he called topcolor--assisted
technicolor (TC2)~\cite{tctwohill}. In TC2, electroweak symmetry breaking is
driven mainly by technicolor interactions strong near $1\,\tev$. Light quark,
lepton, and technipion masses are still generated by ETC. The topcolor
interaction, whose scale is also near $1\,\tev$, generate $\condtbt$ and the
large top--quark mass.\footnote{Three massless Goldstone ``top--pions''
arise from top-quark condensation. The ETC interactions must contribute a
few~GeV to $m_t$ to give the top--pions a mass large enough that $t \ra b
\pi^+_t$ is not a major decay mode.} The scale of ETC interactions still must
be at least several $100\,\tev$ to suppress flavor-changing neutral currents
and, so, the technicolor coupling still must walk. Their marriage neatly
removes the objections that topcolor is unnatural and that technicolor cannot
generate a large top mass. In this scenario, the nonabelian part of topcolor
is an ordinary asymptotically free gauge theory.

Hill's original TC2 scheme assumes separate color $SU(3)$ and weak
hypercharge $U(1)$ gauge interactions for the third and for the first two
generations of quarks and leptons. In the simplest example, the (electroweak
eigenstate) third generation $(t,b)_{L,R}$ transform with the usual quantum
numbers under the topcolor gauge group $\suone \otimes \uone$ while $(u,d)$,
$(c,s)$ transform under a separate group $\sutwo \otimes \utwo$. Leptons of
the third and the first two generations transform in the obvious way to
cancel gauge anomalies. At a scale of order $1\,\tev$, $\suone \otimes \sutwo
\otimes \uone \otimes \utwo$ is dynamically broken to the diagonal subgroup
of ordinary color and weak hypercharge, $SU(3)_C \otimes \uy$. At this
energy, the $\suone \otimes \uone$ couplings are strong while the $\sutwo
\otimes \utwo$ couplings are weak. This breaking gives rise to massive gauge
bosons---a color octet of ``colorons'' $V_8$ and a color singlet $Z'$.

Top, but not bottom, condensation is driven by the fact that the $\suone
\otimes \uone$ interactions are supercritical for top quarks, but subcritical
for bottom.\footnote{A large bottom condensate is not generated by $\suone$
alone because it is broken and its coupling does not grow stronger as one
descends to lower energies.} The difference between top and bottom is caused
by the $\uone$ couplings of $t_R$ and $b_R$. If this TC2 scenario is to be
natural, i.e., there is no fine--tuning of the $\suone$, the $\uone$
couplings {\it cannot} be weak. To avoid large violations of weak isospin in
this and all other TC2 models~\cite{cdt}, right as well as left--handed
members of individual technifermion doublets $T_{L,R} = (T_U, T_D)_{L,R}$
must carry the same $\uone$ quantum numbers, $Y_{1L}$ and $Y_{1R}$,
respectively~\cite{tctwoklee}.

Hill's simplest TC2 model does not how explain how topcolor is broken. Since
natural topcolor requires it to occur near 1~TeV, the most economical cause
is technifermion condensation. In Ref.~\cite{tctwokl}, it was argued that
this can be done for $\suone\otimes\sutwo\ra SU(3)_C$ by arranging that
technifermion doublets $T_1$ and $T_2$ transforming under $\sutc \otimes
\suone \otimes \sutwo$ as $(\Ntc,3,1)$ and $(\Ntc,1,3)$, respectively,
condense with each other as well as themselves, i.e.,
\be\label{eq:tconds} \langle \ol T_{iL} T_{jR} \rangle = -W_{ij} \Delta_T
\quad (i,j = 1,2)\ts,
\ee
where $W$ is a nondiagonal unitary matrix and $\Delta_T$ the technifermion
condensate of $\CO(\LTC^3)$ (see Section~II). The strongly coupled $\uone$
plays a critical role in tilting $W$ away from the identity, which is the
form of the condensate preferred by the color interactions.

The breaking $\uone\otimes\utwo \ra \uy$ is trickier. In order that there is
a well--defined $U(1)_Y$ boson with standard couplings to all quarks and
leptons, this must occur at a somewhat higher scale, several TeV.  Thus, the
$Z'$ boson from this breaking has a mass of several TeV and is {\it strongly}
coupled to technifermions, at least.\footnote{In Ref.~\cite{tctwokl} the
fermions of the first two generations also need to couple to $\uone$. The
limits on these strong couplings and $M_{Z'}$ from precision electroweak
measurements were studied by Chivukula and Terning~\cite{rscjt}. Another
variant of TC2 has all three generations transforming in the same way under
topcolor~\cite{ccs}. This ``flavor--universal topcolor'' has certain
phenomenological advantages (see the second paper of Ref.~\cite{tctwokl} and
Ref.~\cite{ehsbbmix}), but the problems of the strong $\uone$ coupling
afflict it too.} To employ technicolor in this $U(1)$ breaking too,
technifermions $\psi_{L,R}$ belonging to a higher--dimensional $\sutc$
representation are introduced. They condense at higher energy than the
fundamentals $T_{iL,R}$~\cite{multiklee}. The critical reader will note that
this scenario also flirts with unnatural fine tuning because the multi--TeV
$Z'$ plays a critical role in top and bottom quark condensation. Another
pitfall is that the strong $\uone$ coupling may blow up at a Landau
singularity at a relatively low energy~\cite{tctwokl,rsctriv}. To avoid this,
unification of $\uone$ with the nonabelian $G_{ETC}$ must occur at a lower
energy still. Finally, we mention that $V_8$ and $Z'$ exchange induce
$B_d$--$\ol B_d$ mixing that is too large unless $M_{V_8, Z'} \simge
5\,\tev$. This, too, implies a fine--tuning of the TC2 couplings to better
than~1\% (see Ref.~\cite{blt} and Section~II.7). It may be possible to evade
this last constraint in flavor--universal TC2~\cite{ehsbbmix}. This is not a
very satisfactory state of affairs, but that is how things stand for now with
TC2. There are many opportunities for improvement.

A variant of topcolor models is called the ``top seesaw''
mechanism~\cite{seesaw}. Its motivation is to realize the original,
supposedly more economical, top--condensate idea of the Higgs boson as a
fermion--antifermion bound state~\cite{topcondref}. Apart from its
fine--tuning problem, topcolor failed because it implied a top mass of about
250~GeV. In top seesaw models, an electroweak singlet fermion $F$ acquires a
dynamical mass of {\it several} TeV. Through mixing of $F$ with the top
quark, it gives the latter a much smaller mass (the seesaw) and the scalar
$\ol F F$ bound state acquires a component with an electroweak symmetry
breaking vacuum expectation value. We'll say no more about these approaches
here as they are off our main line of technicolor and extended
technicolor. The interested reader should consult the reviews in
Refs.~\cite{rscreview,cthehs} and references therein.

\section*{I.5 Open Problems}

My main goal in this lecture is to attract some bright young people to the
dynamical approach to electroweak and flavor symmetry breaking. Many
challenging problems remain open for study there. This and the next lecture
provide a basis for starting to tackle them. All that's needed now are new
ideas, new data, and good luck. Here are the problems that intrigue and vex
me:

\begin{enumerate}

\item{} First, and most difficult, we need a reasonably realistic model of
extended technicolor, or {\it any other} natural, dynamical description of
flavor. To repeat: This is the hardest problem we face in particle
physics. It deserves much more effort. I believe the difficulty of this
problem and the lack of a ``standard model'' of flavor are what have led to
ETC's being in such disfavor. Experiments will be of great help, possibly
inspiring the right new ideas. Certainly, experiments that will be done in
this decade will rule out, or vindicate, the ideas outlined in these
lectures. That is an exciting prospect!

\item{} More tractable, I think, is the problem of constructing a dynamical
theory of the top--quark mass that is natural, i.e., requires no fine--tuning
of parameters, and has no nearby Landau pole. Like topcolor--assisted
technicolor and top--seesaw models, such a theory is bound to have testable
consequences below 2--3~TeV. So hurry---before the experiments get done!

\item{} Neutrino masses are at least as difficult a problem as the top
mass. In particular, it is a great puzzle how ETC interactions could produce
$m_\nu \simle 10^{-7} m_e$. It seems artificial to have to assume an extra
large ETC mass scale just for the neutrinos. Practically no thought has been
has been given to this problem. Is there some simple way to tinker with the
basic ETC mass--generating mechanism, some way to implement a seesaw
mechanism, or must the whole ETC idea be scrapped? The area is wide open.

\item{} My favorite problem is ``vacuum alignment'' and CP
violation~\cite{rfd,align,vacalign,pascos}; this will be the subject of
Lecture~II. The basic idea is this: Spontaneous chiral symmetry breaking
implies the existence of infinitely many degenerate ground states. These are
manifested by the presence of massless Goldstone bosons (technipions). The
``correct'' ground state, i.e., the one on which consistent chiral
perturbation theory for the technipions is to be carried out, is the one
which minimizes the vacuum expectation value of the explicit chiral symmetry
breaking Hamiltonian $\CH'$ generated by ETC. As Dashen discovered, it is
possible that an $\CH'$ that appears to conserve CP actually violates it in
the correct ground state. This provides a beautiful dynamical mechanism for
the CP violation we observe. Or it could lead to disaster---strong CP
violation, with a neutron electric dipole moment ten orders of magnitude
larger than its upper limit of $0.63\times 10^{-25}\,e$--cm~\cite{pdg}. This
field of research is just beginning in earnest. If the strong--CP problem can
be controlled (and we shall see that there is reason to hope it can be!),
there may be several new sources of CP violation that are accessible to
experiment.

\end{enumerate}

\section*{II. CP VIOLATION IN TECHNICOLOR\footnote{This
lecture is an updated version of my talks at PASCOS 2001 and
La~Thuile~2001 conferences~\cite{pascos,lathuile}.}}

\section*{II.1 Outline}

In this lecture I discuss the dynamical approach to CP violation in
technicolor theories and a few of its consequences. I will cover the
following topics:

\begin{enumerate}

\item{} An elementary introduction to vacuum alignment---in a ferromagnet and
  in QCD.

\item{} A brief description of the strong CP problem.

\item{} Vacuum alignment in technicolor theories and the rational--phase
  solutions for the technifermion--aligning matrices.

\item{} A proposal to solve the strong CP problem without an axion or a
  massless up quark.

\item{} The structure of quark mass and mixing matrices in extended
technicolor (ETC) theories with topcolor--assisted technicolor (TC2). In
particular, realistic Cabibbo--Kobayashi--Maskawa (CKM) matrices are easily
generated.

\item{} Flavor--changing neutral current interactions from extended
technicolor and topcolor. The ETC mass scales that appear in these
interactions are estimated in an appendix to this lecture.

\item{} Results on the $K^0$--$\ol K^0$ CP--violating parameter $\epsilon$
  and the $B_d^0$--$\ol B_d^0$ parameter $\sin(2\beta)$.

\end{enumerate}

\section*{II.2 Introduction to Vacuum Alignment\footnote{Much of this
    section is a modernized version of the discussion in Roger Dashen's
classic 1971 paper~\cite{rfd}. All particle physicists should read that
paper!}}

Consider a simple 3--dimensional ferromagnet with Hamiltonian
\be\label{eq:Hzferro}
\CH_0 = - K \sum_{\langle ij\rangle} {\bf S}_i \cdot {\bf S}_j \qquad (K>0)
\ts,
\ee
where the sum is over nearest--neighbor spins. This Hamiltonian has as its
symmetry group $G = O(3)$, the group of rotations in three dimensions. When
we heat this ferromagnet, the symmetry is manifest because the spins point
every which--way. The ground state $|\Omega\rangle$ of the ferromagnet is
unique, transforming as the spin--zero representation of $O(3)$, so that the
$O(3)$ generators ${\bf J} = \sum_i {\bf S}_i$ annihilate it: ${\bf
J}|\Omega\rangle = 0$.  The ground state's excitations transform as
irreducible representations of $O(3)$.

Now cool the ferromagnet. When it cools below its Curie temperature, $T_C$,
all the spins line up and there is a nonzero magnetization ${\bf M} =
\langle\Omega|{\bf J}|\Omega\rangle$. Now, clearly, the ground state is not
$O(3)$--invariant because not all the generators annihilate it. The symmetry
of $|\Omega\rangle$ is $S = O(2)$, the group of rotations about the axis of
magnetization ${\bf M}$. The symmetry of the Hamiltonian is still $O(3)$, but
the ground state has a lower symmetry. We say that $G=O(3)$ has {\it
spontaneously broken} to $S=O(2)$. Furthermore, the ground state is {\it not
unique}. There are infinitely many $|\Omega\rangle$'s, corresponding to all
the directions in space that ${\bf M}$ can point.\footnote{This degeneracy of
the ground state is manifested by massless excitations, phonons, analogous to
what we call Goldstone bosons.}  We can label these ground states as
$|\Omega(W)\rangle$ where $W$ is an element of $G$ that is not in $S$. This
is called the coset space and written as $G/S = O(3)/O(2)$. All these
``vacua'' have the same energy,
\be\label{eq:vacEzero}
E_0(W) = \langle\Omega(W)|\CH_0|\Omega(W)\rangle \equiv
\langle\Omega|W^{-1}\CH_0 W|\Omega\rangle = E_0(W_0) \ts,
\ee
where $|\Omega\rangle$ is some fixed ``standard vacuum'' corresponding to a
particular $W_0 \in G/S$, say, the one with all spins up along the
$z$--axis. In Eq.~(\ref{eq:vacEzero}) we used the fact that $\CH_0$ is
invariant under all $G$--transformations.

Next turn on a magnetic field in the $z$--direction, ${\bf B} = {\rm B}
\hat{\bf z}$. The Hamiltonian becomes
\be\label{eq:HBferro}
\CH = \CH_0 + \CH' \equiv - K \ts \sum_{\langle ij\rangle} {\bf S}_i
\cdot {\bf S}_j - \mu\sum_i {\bf S}_i \cdot {\bf B} \ts,
\ee
where the magnetic moment $\mu$ is assumed positive for simplicity. The
perturbation $\CH'$ {\it explicitly breaks} the rotational symmetry $O(3)$
down to the particular $O(2)$ corresponding to rotations about the $z$--axis.
Now all the spins will line up pointing along ${\bf B}$ because that
corresponds to the state of lowest energy, i.e.,
\bea\label{eq:vacEB}
E(W_0) &=& \langle\Omega|\CH|\Omega\rangle \equiv
E_0 + \langle\Omega|\CH'|\Omega\rangle\\
 &=& E_0 - \mu {\rm BM} < E(W) {\rm
  \ts\ts for \ts\ts every\ts\ts}  W \neq W_0\ts.\nn
\eea
The vacuum has {\it aligned} with the explicit symmetry breaking perturbation
$\CH'$. 

In more complicated problems, we will not know {\it a priori} which vacuum
``aligns'' with $\CH'$, but the procedure for finding it is clear: we find
the ground state of lowest energy in the presence of $\CH'$. If we fail to do
this, then we generally find pseudoGoldstone bosons (PGBs) with {\it
negative} $M^2$. Indeed, $M^2 < 0$ is the hallmark of working with the wrong
ground state.

Vacuum alignment in a Lorentz--invariant quantum field theory proceeds as
follows~\cite{rfd}: We start with an unperturbed Hamiltonian $\CH_0$ with
symmetry group $G$. The (usually strong) interactions of $\CH_0$ cause $G$ to
spontaneously break to a subgroup $S$, the symmetry group of the degenerate
ground states of $\CH_0$. The infinitely many ground states,
$|\Omega(W)\rangle$, are parameterized by $W \in G/S$. Associated with this
spontaneous symmetry breaking are $n$ massless Goldstone bosons, one for each
of the $N$ continuous symmetry generators $Q_a$ ($a=1,2,\dots,N$) of
$G/S$.\footnote{Strictly speaking, the definition of $S$ and the $Q_a$
corresponds to the choice of $W$ defining $|\Omega(W)\rangle$, but that won't
concern us here.} The degeneracy of these vacua may be partially or wholly
lifted by the explicit symmetry breaking perturbation, $\CH'$. In practical
cases, it is necessary that matrix elements of $\CH'$ are small compared to
those of $\CH_0$. Then we can find the correct ground state(s) upon which to
carry out a perturbation expansion in $\CH'$ by minimizing the vacuum energy,
\be\label{eq:vacEeq} 
E(W) = \langle\Omega(W)|\CH|\Omega(W)\rangle \equiv
\langle\Omega|W^{-1}\CH' \ts W|\Omega\rangle + \ts {\rm constant}\ts,
\ee
over the transformations $W\in G/S$. The standard vacuum $|\Omega\rangle$ is
chosen to give simple forms for the vacuum expectation values of operators,
usually fermion bilinears and quadrilinears.

Finally, let us write $Q_a = \int d^3x j_0^a(x)$ for the charge generating a
transformation in $G/S$. Here, $j_0^a$ is the time component of the current
whose divergence is the equal--time commutator
\be\label{eq:DJ}
\partial^\mu j_\mu^a(x) = i \ts [\CH'(x), Q_a(x_0)] \ts.
\ee
Then, Dashen showed that, for $W_0 \in G/S$ which minimizes $E(W)$, the
condition for an extremum of the vacuum energy is
\be\label{eq:first}
\langle\Omega|[Q_a, W_0^{-1} \CH' W_0 ]|\Omega\rangle = 0 \qquad
(a=1,\dots,N) \ts.
\ee
This is trivial if $Q_a$ is a generator of the symmetry group of
$|\Omega\rangle$, for then $Q_a|\Omega\rangle = 0$. Furthermore, the condition
for $E(W_0)$ to be a minimum is that the PGB mass--squared matrix
\be\label{eq:second}
(f^2_\pi M^2_\pi)_{ab} = i^2 \langle\Omega|[Q_a,[Q_b, W_0^{-1}\CH'
  W_0]] |\Omega\rangle \ts,
\ee
is positive--semidefinite. Here, $f_\pi$ is the decay constant matrix of the
PGBs. In most cases of interest to us, $f_\pi$ is a single constant. Note
that Eq.~(\ref{eq:first}) plus the Jacobi identity imply that this is a
symmetric matrix.

Consider vacuum alignment for QCD. The unperturbed Lagrangian for three {\it
  massless} quarks $u,d,s$ is
\be\label{eq:LQCD} \CL_0 = -\fourth \ts F_{\mu\nu}^A F^{A,\mu\nu} +
\sum_{j=1}^3 i\ts \ol q_j \gamma^\mu \CD_\mu q_j \ts,
\ee
where $F_{\mu\nu}^A$ is the gluon field strength and $\CD_\mu = \partial_\mu
- ig_C t_A G^A_\mu$ is the quarks' covariant derivative (QCD coupling $g_C$
and $SU(3)_C$ generator $t_A$ for $A=1,\dots,8$). The Hamiltonian $\CH_0$ is
easily constructed from $\CL_0$, but it's not as pretty. When the quarks are
massless, $\CH_0$ is invariant under separate (global) $SU(3)$
transformations on the left and right--handed quark fields. This Hamiltonian
is also invariant under the discrete CP symmetry. Because the quarks are
massless, we are free to work with vacua whose instanton angle $\theta_{QCD}
= 0$, so there is no CP--violating $F_{\mu\nu}^A \tilde F^{A,\mu\nu}$
interaction in $\CL_0$; see Section~II.3 and Ref.~\cite{CPreview} for a
discussion of this. So, the symmetry group of $\CH_0$ is\footnote{As
discussed in Section~II.3, instantons strongly break the axial $U(1)$ symmetry
so that, ignoring a trivial vectorial $U(1)$, the continuous flavor symmetry
group of $\CH_0$ is $SU(3)_L \otimes SU(3)_R$ instead of $U(3)_L \otimes
U(3)_R$.}
\be\label{eq:GQCD}
G_f = SU(3)_L \otimes SU(3)_R \otimes {\rm CP} \ts.
\ee
The $SU(3)_L \otimes SU(3)_R$ is called the quarks' {\it chiral} symmetry
group.

When QCD interactions become strong at energies of order a few hundred MeV,
quark condensates form and $G$ is spontaneously broken to $S_f = SU(3)
\otimes {\rm CP}$:
\be\label{eq:qcond}
\langle\Omega(W)|\ol q_{Rj} q_{Li}|\Omega(W)\rangle = - \half \Delta_q
W_{ij} \ts.
\ee
Here, $W = W_L W_R^\dag \in G_f/S_f \sim SU(3)$; i.e., $W$ is a unitary {\it
unimodular} $SU(3)$ matrix parameterizing the ground state
$|\Omega(W)\rangle$. We take our standard vacuum to be the one with diagonal
condensates, i.e., with $W = 1$ in Eq.~(\ref{eq:qcond}) and $S=SU(3)_V$, the
diagonal subgroup of $SU(3)_L \otimes SU(3)_R$ generated by vector charges
$Q_a$. These charges annihilate the standard vacuum: $Q_a|\Omega\rangle =
0$. The eight Goldstone bosons are the familiar $\pi_a = \pi^{\pm,0},
K^{\pm}, K^0, \ol K^0, \eta$. They correspond to the axial charges
\bea\label{eq:axial}
&&Q_{5a} = \int d^3x \ts j^a_{50}(x) \ts;\nn\\
&&\langle\Omega|j^a_{5\mu} |\pi_b(p)\rangle = i \delta_{ab} f_\pi p_\mu \ts.
\eea
The condensate $\Delta_q \simeq 4\pi f^3_\pi$, where $f_\pi = 93\,\mev$.

Now for the fun! Quarks aren't massless; their chiral $SU(3)_L \otimes
SU(3)_R$ symmetry is explicitly broken (to $U(1)^3$) by
\be\label{eq:CHprime}
\CH' = \ol q_R M_q q_L + \ol q_L M_q^\dag q_R \equiv \ol q M_q q
\ts,
\ee
where the (assumed) real quark mass matrix is
\be\label{eq:Mq}
 M_q = M_q^\dag =
    \pm \left(\ba{ccc}
      m_u & 0 & 0\\
      0 & m_d & 0\\
      0 & 0 & m_s\\
      \ea\right)
\ee
All elements $m_i$ of the quark mass matrix are assumed real and positive and
small compared to the QCD energy scale, approximately $4\pi f_\pi$. Then,
$\CH = \CH_0 + \CH'$ conserves CP. The interesting physics comes from the
sign in front of the quark mass matrix.

To simplify the calculations, we follow Dashen and assume that $\CH'$ is
$SU(2) \otimes U(1)$ invariant, where the $SU(2)$ is ordinary vectorial
isospin:
\be\label{eq:simlify}
m_u = m_d = \delta m_s\ts, \qquad {\rm with} \ts\ts \delta \ge 0\ts.
\ee
Then, for $W\in SU(3)$, we have to minimize
\bea\label{eq:EW}
E(W=W_L W_R^\dag) &=&
\langle\Omega|\ol q_R (W_R^\dag M_q W_L) q_L
            + \ol q_L (W_L^\dag M_q W_R) q_R |\Omega\rangle \nn\\
&=& -\half \Delta_q {\rm Tr}(M_q W + M_q^\dag W^\dag) \equiv
E(W^*) \ts.
\eea
The last equality is a consequence of the reality of the quark mass matrix,
i.e., the CP--invariance of $\CH'$. It tells us that, if $W_0$ minimizes
$E(W)$, then so does $W_0^*$. And, if $W_0$ is complex, then the CP symmetry
is spontaneously broken:\footnote{Because CP is a discrete symmetry, no
Goldstone boson is associated with its spontaneous breaking.} two different
vacua, $|\Omega(W_0)\rangle$ and $|\Omega(W_0^*)\rangle$, correspond to
degenerate minima of $E(W)$ and we must pick one on which to build our
Hilbert space and carry out the perturbation expansion. Now, I present the
vacuum alignment problem as two exercises for the reader:

\nin {\underbar{Exercise 1:}} Show that 
\be\label{eq:Mqplus}
 M_q =
    + m_s\left(\ba{ccc}
      \delta & 0 & 0\\
      0 & \delta & 0\\
      0 & 0 & 1\\
      \ea\right) \quad \Longrightarrow \quad W_0 =
\left(\ba{ccc}
      1 & 0 & 0\\
      0 & 1 & 0\\
      0 & 0 & 1\\
      \ea\right)  \ts\ts {\rm for} \ts\ts \delta \ge 0 \ts;
\ee
and
\bea\label{eq:Mqminus}
&& M_q =
    - m_s\left(\ba{ccc}
      \delta & 0 & 0\\
      0 & \delta & 0\\
      0 & 0 & 1\\
      \ea\right)\nn\\
&&\Longrightarrow \ts\ts W_0 = 
\left(\ba{ccc}
      -1 & 0 & 0\\
      0 & -1 & 0\\
      0 & 0 & 1\\
      \ea\right)  \ts\ts {\rm for} \ts\ts \delta \ge 2 \ts; \\
&&\Longrightarrow \ts\ts W_0 = 
\left(\ba{ccc}
      -\half\delta \pm i\eta & 0 & 0\\
      0 & -\half\delta \pm i\eta & 0\\
      0 & 0 & -1 +\half\delta \pm i\delta\eta\\
      \ea\right)  \ts\ts {\rm for} \ts\ts 0 \le \delta \le 2 \ts,\nn\\
\eea
where $\eta = \sqrt{1-\fourth\delta^2}$ and the $\pm$ sign corresponds to the
two degenerate vacua. That is, for the minus sign and $0 \le \delta \le 2$,
the mass perturbation is
\bea\label{eq:CHCP}
\CH'(W_0) &\equiv& 
\ol q_R \left(W_{0R}^\dag M_q W_{0L}\right) q_L
            + \ol q_L \left(W_{0L}^\dag M_q^\dag W_{0R}\right) q_R \\ \nn \\
&=& \ol q \left(\ba{ccc}
      \half \delta^2 \ts m_s & 0 & 0\\
      0 & \half \delta^2 \ts m_s & 0\\
      0 & 0 & (1-\half \delta^2) \ts m_s\\
      \ea\right)q \mp i\delta\eta \ts m_s \ts \ol q\gamma_5 q \ts.\nn
\eea
and it is CP--violating! As Dashen said back in 1971: ``It would be nice to
think that the CP violation observed in weak interactions comes about through
a phenomenon like that described above. Clearly, Eq.~(\ref{eq:CHCP}) has
nothing to do with the observed CP violation. Equation~(\ref{eq:CHCP}) gives
CP violation whose strength is on the order of $SU(3)\otimes SU(3)$ breaking
which is orders of magnitude too strong. The interesting thing about
Eq.~(\ref{eq:CHCP}) is that it shows that spontaneous CP violation can really
occur and that, in certain circumstances, one can actually predict when it
will occur. One simply has to take an otherwise harmless looking $\CH'$ of
the class given by Eq.~(\ref{eq:CHprime}) [with the minus sign], choose
$\delta < 2$ and out comes CP violation. It is a very interesting question
whether some more complex or simply more clever theory could give spontaneous
CP--violating effects of the magnitude observed in weak interactions.'' This
is precisely what we propose arises from vacuum alignment in
technicolor.\footnote{Spontaneous CP violation at the TeV scale gives rise to
cosmic domain walls whose energy density can over--close the universe. It is
an open question whether there is some mechanism for getting rid of the
domain walls in our technicolor scenario.}

\nin {\underbar{Exercise 2:}} Calculate the PGB mass--squared matrix to show
that it is given by
\bea\label{eq:Mpisq}
f_\pi^2 \bigl(M^2_\pi\bigr)_{ab} &\equiv& i^2
\langle\Omega|[Q_{5a},[Q_{5b}, \CH'(W_0)]] |\Omega\rangle \nn\\
&=& {\rm Tr}\bigl[\bigr\{t_a,\bigl\{t_b,M_q W_0\bigr\}\bigr\} +
{\rm h.c.} \bigr] \Delta_q \ts,
\eea
where $t_a = \lambda_a/2$, the Gell-Mann matrices for $SU(3)$. Evaluate
$(M^2_\pi)_{ab}$ and find its eigenvalues for both signs of $M_q$ and for
$\delta \ge 2$ and $0 \le \delta \le 2$. What happens to the PGB masses at
the boundaries $M_q \ra 0^+$ and, for $M_q$ negative, at $\delta \ra 2$?
Explain!

\section*{II.3 The Strong CP Problem}

In Section~II.2 we made passing reference to instantons, the angle
$\theta_{QCD}$, and CP violation induced by the gluon term $F\cdot\tilde
F$. Here I'll briefly describe what this is all about; see
Ref.~\cite{CPreview} for more details.

Nonabelian gauge theories such as QCD and technicolor (and even electroweak
$SU(2)$) have topologically nontrivial gauge field configurations for which
the integral $\int d^4x F\cdot\tilde F$ does not vanish even though
$F\cdot\tilde F$ can be written as a total derivative. In fact,
\be\label{eq:winding}
{g_C^2\over{32\pi^2}}\int d^4x \ts F_{\mu\nu}^A \tilde F^{A,\mu\nu} = n
\ts,
\ee
where the integer $n=0,\pm1,\pm2,\dots$ is called the ``winding number'' of
the gauge field. For $n\neq 0$, these configurations, called instantons, fall
off sufficiently slowly at large distances that the surface integral at
infinity does not vanish. The net result of the existence of instantons is
that one must choose the ``$\theta$--vacua'' as ground states of the
nonabelian gauge theory. These are defined by $|\Omega(\theta)\rangle =
\sum_n \exp{(i \ts n\theta)} |\Omega(n)\rangle$, where $|\Omega(n)\rangle$ is
the ground state in the presence of an instanton with winding number~$n$. The
fact that we are in a $\theta$--vacuum may be expressed in the Lagrangian by
adding the term
\be\label{eq:LFFdual}
\CL_{\theta} = {g_C^2\over{32\pi^2}} \theta F_{\mu\nu}^A \tilde
F^{A,\mu\nu} \ts.
\ee
If the matter and gauge fields in the Lagrangian are defined to transform
under CP and T in the usual way, this interaction is odd under those
symmetries and violates them.

Now consider QCD with six quarks whose Lagrangian is $\CL_0$ in
Eq.~(\ref{eq:LQCD}) plus $\CL_\theta$ and a mass term like $\CH'$ above,
\be\label{eq:CLm}
\CL_m = -\sum_{j=1}^6 m_j \ol q_j q_j \ts.
\ee
The charge of the axial vector current $j_{5\mu} = \sum_j \ol
q_j\gamma_\mu\gamma_5 q_j \equiv j_{R\mu} - j_{L\mu}$ generates the $U(1)_A$
transformations $q_{Rj} \ra \exp{(i\alpha)} q_{Rj}$ and $q_{Lj} \ra
\exp{(-i\alpha)} q_{Lj}$. But these transformations are not a symmetry of
QCD, because the current is not conserved:
\be\label{eq:axialdiv}
\partial^\mu j_{5\mu} = 2i\sum_{j=1}^6 m_j \ol q_j \gamma_5 q_j +
{6g_C^2 \over{16\pi^2}} F_{\mu\nu}^A \tilde F^{A,\mu\nu} \ts.
\ee
If we make a $U(1)_A$ rotation by angle $\alpha$, the QCD Lagrangian changes
as
\bea\label{eq:deltaL} &&\CL_{QCD} \equiv \CL_0 + \CL_m + \CL_{\theta_{QCD}}
\\
&&\ts\ts\ra \CL_0 + \sum_{j=1}^6 m_j \bigl[e^{2i\alpha} \ts \ol q_{Lj} q_{Rj}
+ e^{-2i\alpha} \ts \ol q_{Rj} q_{Lj}\bigr] + {g_C^2 \over{32\pi^2}}
(\theta_{QCD} + 12\alpha)F_{\mu\nu}^A \tilde F^{A,\mu\nu} \ts. \nn
\eea
We see that a $U(1)_A$ rotation by $\alpha$ changes $\theta_{QCD}$ by
$6\alpha$, the effect showing up in the quark mass terms. This is a canonical
transformation; the theories based on the two Lagrangians are completely
equivalent.

If all the quarks are massless, more precisely, if {\it all} mass terms in
$\CL_{QCD}$ vanish, the $U(1)_A$ current is still not conserved. The
$F\cdot\tilde F$ term is called the axial current's anomalous divergence, or
the axial anomaly for short. Its matrix elements are large,
$\CO(\Lambda_{QCD}^4)$, and so this current is not even approximately
conserved. As noted earlier, this means that $U(1)_A$ is not a good symmetry
of QCD, and there is (for three light quarks) no ninth Goldstone boson
associated with spontaneous chiral symmetry breaking by the quark condensates
$\langle\ol q q\rangle$. This is why, when we carried out QCD vacuum
alignment in the previous section, we restricted ourselves to {\it
unimodular} $SU(3)$ matrices in $G_f/S_f$. It was this constraint of
unimodularity that led to CP violation in the example you worked out. For the
minus sign and $\delta < 2$, the vacuum energy has a lower minimum if you use
a nonunimodular $U(3)$ matrix, but that is not allowed because of the
instantons. Another way to say the same thing is that, in this case, the
quark mass matrix is brought to real, positive, diagonal form by a $U(3)$,
not $SU(3)$, matrices.

On the other hand, if any of the quarks were massless, we could rotate
$\theta_{QCD}$ to zero by choosing $\alpha = -\theta_{QCD}/12$. More
precisely, in this case $\theta_{QCD}$ is unobservable, and there is no
CP--violation associated with the QCD instantons. In the standard model, the
quark mass terms aren't zero and, so, there can be instanton--induced CP
violation. The observable measure of this is the angle
\be\label{eq:thqbar}
\ol \theta_q = -\theta_{QCD} + \arg\det{M_q}\ts;
\ee
this is unchanged by $U(1)_A$ rotations. There is every reason to expect
$\ol \theta_q = \CO(1)$ in the standard model, at the very least because the
phases of Yukawa couplings of the Higgs boson to quarks are arbitrary. That
is a catastrophe, because this CP violation implies a neutron electric dipole
moment of~\cite{baluni}
\be\label{eq:edm}
d_N \simeq {e m_u\sin\ol\theta_q\over{M^2_N}} \simeq 10^{-16} \sin\ol\theta_q
\ts {\rm e-cm}\ts.
\ee
This is almost ten orders of magnitude larger than the upper limit
$0.63\times 10^{-25}\,e$--cm.\cite{pdg}. We need $|\ol\theta_q| \simle
10^{-10}$.

This is the strong--CP problem. Many proposals have been made over the past
24~years for solving it~\cite{CPreview}. The most elegant is the
Peccei--Quinn $U(1)$ symmetry which relaxes $\ol \theta_q$ to zero in tree
approximation (it remains suitably tiny in higher orders) by extending the
$U(1)_A$ to two doublets of Higgs fields. This symmetry is spontaneously
broken by the Higgs vacuum expectation values. The resulting Goldstone boson
is the axion. Its mass is $M_a \simeq f_\pi M_\pi/f_a$ and its coupling to a
fermion $f$ with hard mass $m_f$ is approximately $m_f/f_a$, where $f_a
\simle 246\,\gev$ is the axion's decay constant. This is ruled out entirely!
Other popular mechanisms include the so-called invisible axion, the
Barr--Nelson mechanism, and a massless up quark. These proposals are also
ruled out experimentally or rather contrived and clumsy, though still
allowed. In this lecture I will describe another possibility, based on vacuum
alignment in technicolor. Originally, it was thought that technicolor
theories, without bare mass terms for the fermions, would trivially solve the
strong--CP problem by making $\ol \theta_q$ unobservable. This is
wrong. These theories still have explicit chiral symmetry breaking, as they
must, via ETC interactions. While the QCD and TC instanton angles can be
rotated to zero, their effects are still observable through $\ol\theta_q$.

\section*{II.4 Vacuum Alignment in the Technifermion Sector}

The idea that the observed weak CP violation arises from vacuum alignment
finds its natural home in technicolor~\cite{tc,frascati,rscreview,cthehs}
where large groups of flavor/chiral symmetries are spontaneously broken by
strong dynamics and explicitly broken by extended technicolor
(ETC)~\cite{etceekl}. Furthermore, the perturbation $\CH'$ generated by
exchange of ETC gauge bosons generally is {\it naively} CP--conserving if CP
is unbroken above the technicolor energy scale. Motivated by this, Eichten,
Preskill and I proposed in 1979 that CP violation occurs spontaneously in
theories of dynamical electroweak symmetry breaking~\cite{align}. Our goal,
unrealized at the time, was to solve QCD's strong--CP problem {\it without}
invoking a Peccei--Quinn symmetry and its accompanying axion or requiring
that the up quark is massless.

This problem was taken up again a few years ago with Eichten and
Rador~\cite{vacalign}. We studied the first important step in reaching this
goal: vacuum alignment in the technifermion sector. Once this is carried out,
alignment in the quark sector is determined by the technifermion aligning
matrices and the ETC couplings of quarks to technifermions. This will be
described in more detail below.

We considered models in which a single kind of technifermion interacts with
quarks via ETC interactions. Leptons are ignored for now. We assume $N$
technifermion doublets $T_{L,R\ts I} = (\CU_{L,R\ts I}, \ts \CD_{L,R\ts I})$,
$I = 1,2,\dots,N$, all transforming according to the fundamental
representation of the technicolor gauge group $SU(N_{TC})$. There are three
generations of $SU(3)_C$ triplet quarks $q_{L,R\ts i} = (u_{L,R\ts i}, \ts
d_{L,R\ts i})$, $i = 1,2,3$. The left--handed fermions are electroweak
$SU(2)$ doublets and the right--handed ones are singlets. Here and below, we
exhibit only flavor, not technicolor and color, indices.

The technifermions are assumed for simplicity to be ordinary color--singlets,
so the chiral flavor group of our model is $G_f = \left[SU(2N)_{L} \otimes
SU(2N)_{R}\right] \otimes\left[SU(6)_{L} \otimes
SU(6)_{R}\right]$.\footnote{The fact that heavy quark chiral symmetries
cannot be treated by chiral perturbative methods will be addressed below. We
have excluded anomalous $U_A(1)$'s strongly broken by TC and color instanton
effects. Therefore, alignment matrices must be unimodular.} When the TC and
QCD couplings reach their required critical values, these symmetries are
spontaneously broken to $S_f = SU(2N) \otimes SU(6)$. We adopt as the
standard vacuum the ground state $|\Omega\rangle$ whose symmetry group is the
the vectorial $SU(2N)_V \otimes SU(6)_V$, Then the fermion bilinear
condensates given by
\bea\label{eq:standard}
\langle \Omega |\ol \CU_{LI} \CU_{RJ}|\Omega \rangle &=&
\langle \Omega |\ol \CD_{LI} \CD_{RJ}|\Omega \rangle = -\delta_{IJ} \Delta_T
\nn\\
\langle \Omega |\ol u_{Li} u_{Rj}|\Omega \rangle &=&
\langle \Omega |\ol d_{Li} d_{Rj}|\Omega \rangle = -\delta_{ij} \Delta_q \ts.
\eea
Here, $\Delta_T \simeq 4\pi F_T^3$ and $\Delta_q \simeq 4\pi f_\pi^3$ where
$F_T = 246\,\gev/\sqrt{N}$ is the technipion decay constant.

All of the $G_f$ symmetries except for the gauged electroweak $SU(2) \otimes
U(1)$ are explicitly broken by ETC interactions. In the absence of a concrete
model, we write the interactions broken at the scale~\footnote{See the
Appendix for estimates of $\METC/\getc$ in walking technicolor.} $\METC/\getc
\sim 10^2$--$10^4\,\tev$ in the phenomenological four-fermion form (sum over
repeated indices)~\footnote{We assume that ETC interactions commute with
electroweak $SU(2)$, though not with $U(1)$ nor color $SU(3)$. All fields in
Eq.~(2) are electroweak, not mass, eigenstates.}
\bea\label{eq:Hetc}
\CH' &\equiv& \CH'_{TT} + \CH'_{Tq} + \CH'_{qq} \nn\\
&=& \Lambda^{TT}_{IJKL} \ts \ol{T}_{LI}\gamma^{\mu}T_{LJ}
\ts \ol{T}_{RK}\gamma_{\mu}T_{RL} \nn 
+ \Lambda^{Tq}_{IijJ} \ts \ol{T}_{LI}\gamma^{\mu}q_{Li}
\ts \ol{q}_{Rj}\gamma_{\mu}T_{RJ} + {\rm h.c.}\\
&+& \Lambda^{qq}_{ijkl} \ts \ol{q}_{Li}\gamma_{\mu}q_{Lj}
\ts \ol{q}_{Rk}\gamma_{\mu}q_{Rl} \ts.
\eea
Here, the fields $T_{L,R\ts I}$ and $q_{L,R\ts i}$ stand for all $2N$
technifermions and six quarks, respectively. The $\Lambda$ coefficients are
$\CO(g^2_{ETC}/M^2_{ETC})$ times mixing factors for these bosons and group
theoretical factors for the broken generators of ETC. The $\Lambda$'s may
have either sign. In all calculations, we must choose the $\Lambda$'s to
avoid unwanted Goldstone bosons. Hermiticity of $\CH'$ requires
\be\label{eq:herm}
(\Lambda^{TT}_{IJKL})^* = \Lambda^{TT}_{JILK} \ts, \qquad
(\Lambda^{Tq}_{IijJ})^* = \Lambda^{Tq}_{iIJj} \ts, \qquad
(\Lambda^{qq}_{ijkl})^* = \Lambda^{qq}_{jilk} \ts.
\ee
Assuming, for simplicity, that color and technicolor are embedded in a simple
nonabelian ETC group, the instanton angles $\theta_{TC}$ and $\theta_{QCD}$
are equal. Without loss of generality, we may work in vacua in which they are
zero. Then $\ol \theta_q = \arg\det(M_q)$. The assumption of time--reversal
invariance for this theory before any potential breaking via vacuum alignment
then means that all the $\Lambda$'s are {\it real} and so
$\Lambda^{TT}_{IJKL} = \Lambda^{TT}_{JILK}$, etc.

Having chosen a standard $|\Omega\rangle$, vacuum alignment proceeds by
minimizing the expectation value of the $G_f/S_f$--rotated Hamiltonian. This
is obtained by making the transformations $T_{L,R} \ra W_{L,R} \ts
T_{L,R}$ and $q_{L,R} \ra Q_{L,R} \ts q_{L,R}$, where $W_{L,R} \in
SU(2N)_{L,R}$ and $Q_{L,R} \in SU(6)_{L,R}$:
\bea\label{eq:HW}
\CH'(W,Q) &=& \CH'_{TT}(W_L,W_R) +  \CH'_{Tq}(W,Q) +
\CH'_{qq}(Q_L,Q_R) \\
&=& \Lambda^{TT}_{IJKL} \ts \ol{T}_{LI'} W_{L\ts I'I}^\dag
\gamma^{\mu}W_{L\ts JJ'}T_{LJ'} \ts \ol{T}_{RK'} W_{R\ts K'K}^\dag
\gamma^{\mu}W_{R\ts LL'}T_{RL'} + \cdots \ts.\nn
\eea
So long as vacuum alignment preserves electric charge conservation, the
alignment matrices will be block--diagonal
\bea\label{block}
W_{L,R} =  \left(\ba{cc} W^U & 0 \\ 0 & W^D \ea\right)_{L,R} \ts; \qquad
Q_{L,R} =  \left(\ba{cc} U & 0 \\ 0 & D \ea\right)_{L,R} \ts.
\eea
Since $T$ and $q$ transform according to complex representations of their
respective color groups, the four--fermion condensates in the
$S_f$--invariant $|\Omega\rangle$ have the form
\bea\label{eq:conds}
\langle\Omega|\ol{T}_{LI}\gamma^{\mu}T_{LJ}
\ts \ol{T}_{RK}\gamma_{\mu}T_{RL}|\Omega\rangle &=& -\Delta_{TT}
\delta_{IL}\delta_{JK} \ts, \nonumber \\
\langle\Omega| \ol{T}_{LI}\gamma^{\mu}q_{Li}
\ts \ol{q}_{Rj}\gamma_{\mu}T_{RJ} |\Omega\rangle &=&
-\Delta_{Tq}\delta_{IJ}\delta_{ij} \ts, \\
\langle\Omega|\ol{q}_{Li}\gamma^{\mu}q_{Lj}
\ts \ol{q}_{Rk}\gamma_{\mu}q_{Rl} |\Omega\rangle &=&
-\Delta_{qq}\delta_{il}\delta_{jk} \ts. \nonumber 
\eea
The condensates are positive, renormalized at $\METC$ and, in the limits of a
large number of technicolors and colors, $\Ntc$ and $N_C$, they are given by
\bea\label{eq:largeNconds}
\Delta_{TT} &\simeq& (\Delta_T(\METC))^2 \nn\\
\Delta_{Tq} &\simeq& \Delta_T(\METC) \ts \Delta_q(\METC)\\
\Delta_{qq} &\simeq& (\Delta_q(\METC))^2\ts.\nn
\eea
In walking technicolor~\cite{wtc} (see the Appendix)
\be\label{eq:tccond}
\Delta_T(\METC) \simle(\METC/\Lambda_{TC}) \ts \Delta_T(\Lambda_{TC}) =
10^2-10^4 \times \Delta_T(\Lambda_{TC})\ts.
\ee
In QCD, however,
\be\label{qcdcond}
\Delta_q(\METC) \simeq (\log(\METC/\Lambda_{QCD}))^{\gamma_m} \ts
\Delta_q(\Lambda_{QCD}) \simeq \Delta_q(\Lambda_{QCD})\ts,
\ee
where, from Eq.~(\ref{eq:nonpert}), the anomalous dimension of $\ol q q$ is
$\gamma_m \simeq 2\alpha_C/\pi$ for color $SU(3)_C$. This implies that the
ratio
\be\label{eq:ratio}
r = {\Delta_{Tq}(\METC) \over{\Delta_{TT}(\METC)}} \simeq
{\Delta_{qq}(\METC) \over{\Delta_{Tq}(\METC)}} \simeq
{\Lambda_{TC}\over{\METC}} \left({f_\pi\over{F_T}}\right)^3 \simle 10^{-11} 
\ee
for $F_T \simeq 100\,\gev$. This ratio is $10^2$--$10^4$ times smaller than
it is in a technicolor theory in which the coupling does not walk.

With these condensates, the vacuum energy is a function only of $W = W_L \ts
W_R^\dag$ and $Q = Q_L \ts Q_R^\dag$, elements of the $G_f/S_f$:
\bea\label{eq:vacE}
& &E(W,Q) = E_{TT}(W) + E_{Tq}(W,Q) + E_{qq}(Q) \\
& & \ts\ts = -\Lambda^{TT}_{IJKL} \ts W_{JK} \ts W^\dag_{LI} \ts \Delta_{TT}
       -\left(\Lambda^{Tq}_{IijJ} \ts Q_{ij} \ts W^\dag_{JI} + {\rm c.c.}
         \right) \Delta_{Tq} 
       -\Lambda^{qq}_{ijkl} \ts Q_{jk} \ts Q^\dag_{li} \ts \Delta_{qq} \nn \\
& & \ts\ts = -\Lambda^{TT}_{IJKL} \ts W_{JK} \ts W^\dag_{LI} \ts \Delta_{TT}
+\CO(10^{-11}) \ts.\nn 
\eea
Note that time--reversal invariance of the unrotated Hamiltonian $\CH'$
implies that $E(W,Q) = E(W^*,Q^*)$. Hence, spontaneous CP violation occurs if
the solutions $W_0$, $Q_0$ to the minimization problem are not real up to an
overall phase in $\CZ_N$.

The last line of Eq.~(\ref{eq:vacE}) makes clear that we should first
minimize energy $E_{TT}$ in the technifermion sector. This determines $W_0$,
and as we shall see, $\ol\theta_q$, up to corrections of $\CO(10^{-11})$ from
the quark sector.\footnote{Two other sorts of corrections need to be
studied. The first are higher--order ETC and electroweak contributions to
$E_{TT}$. The electroweak ones are naively $\CO(10^{-7})$, much too large for
$\ol\theta_q$. The second are due to $\ol T t \ol t T$ terms in $E_{Tq}$
which may be important if the top condensate is large. I thank J.~Donoghue
and S.~L.~Glashow for emphasizing the potential importance of these
corrections.} This result is then fed into $E_{Tq}$ to determine $Q_0$---and
the nature of {\it weak} CP violation in the quark sector---up to corrections
which are also $\CO(10^{-11})$.

In Ref.~\cite{vacalign}, it was shown that just three possibilities naturally
occur for the phases in $W$. (We drop its subscript ``0'' from now on.) Let
us write $W_{IJ} = |W_{IJ}| \exp{(i\phi_{IJ})}$. Consider an individual term
$-\Lambda^{TT}_{IJKL} \ts W_{JK} \ts W^\dag_{LI} \ts \Delta_{TT}$ in $E_{TT}$.
If $\Lambda^{TT}_{IJKL} > 0$, this term is least if $\phi_{IL}
= \phi_{JK}$; if $\Lambda^{TT}_{IJKL} < 0$, it is least if $\phi_{IL} =
\phi_{JK} \pm \pi$. We say that $\Lambda^{TT}_{IJKL} \neq 0$ links
$\phi_{IL}$ and $\phi_{JK}$, and tends to align (or antialign) them. Of
course, the constraints of unitarity may partially or wholly frustrate this
alignment. The three possibilities we found for the phases are:

\begin{enumerate}

\item{} The phases are all unequal, irrational multiples of $\pi$ that are
random except for the constraints of unitarity and unimodularity.

\item{} All of the phases may be equal to the same integer multiple of
$2\pi/N$ (mod~$\pi$). This occurs when all phases are linked and aligned, and
the value $2\pi/N$ is a consequence of {\it unimodularity}.\footnote{Because
$W$ is block diagonal, $E_{TT}$ factorizes into two terms, $E_{UU} + E_{DD}$,
in which $W_U$ and $W_D$ may each be taken unimodular. Therefore, totally
aligned phases are multiples of $2\pi/N$, not $\pi/N$.} In this case we say
that the phases are ``rational''.

\item{} Several groups of phases may be linked among themselves and the
phases only partially aligned. In this case, their values are various
rational multiples of $\pi/N'$ for one or more integers $N'$ from~1 to~$N$.

\end{enumerate}

\nin We stress that, as far as we know, rational phases occur naturally only
in ETC theories. They are a consequence of $E_{TT}$ being quadratic, not
linear, in $W$ and of the instanton induced unimodularity constraints. With
these three outcomes in hand, we proceed to investigate the strong CP
violation problem of quarks.

\section*{II.5 A Dynamical Solution to the Strong--CP Problem}

To recapitulate: There are two kinds of CP violation in the quark
sector. Weak CP violation enters the standard weak interactions through the
CKM phase $\delta_{13}$ and, for us, in the ETC and TC2 interactions through
phases in the quark alignment alignment matrices $U_{L,R}$ and $D_{L,R}$
discussed in Section~II.6. Strong CP violation, which can produce electric
dipole moments $10^{10}$ times larger than the experimental bound, is a
consequence of instantons~\cite{CPreview}. No discussion of the origin of CP
violation is complete which does not eliminate strong CP violation. Resolving
the strong CP problem amounts to making $\ol \theta_q = \arg\det(M_q) \simle
10^{-10}$ (in a basis with instanton angle $\theta_{QCD} = 0$). Here, $M_q$
is the ``hard'' or ``current algebra'' mass matrix of the quarks, running
only logarithmically with energy up to the ETC scale (see Section~I.2).

The element $(\CM_q)_{ij}$ of the ``primordial'' quark mass matrix, the
coefficient of the bilinear $\ol q'_{Ri} q'_{Lj}$ of quark {\it electroweak}
eigenstates, is generated by ETC interactions and is given by\footnote{The
matrix element $\CM_{tt}$ arises almost entirely from the TC2--induced
condensation of top quarks. We assume that $\langle \ol t t \rangle$ and
$\CM_{tt}$ are real in the basis with $\theta_{QCD} = 0$. Since technicolor,
color, and topcolor groups are embedded in ETC, all CP--conserving
condensates are real in this basis.}
\be\label{eq:primordial}
 (\CM_q)_{ij} = \sum_{I,J} \Lambda^{Tq}_{IijJ} \ts W^\dag_{JI} \ts
 \Delta_T(\METC)  \qquad (q,T = u,U \ts\ts {\rm or} \ts\ts d,D)\ts.
\ee
The $\Lambda^{Tq}_{IijJ}$ are real ETC couplings of order
$(10^2$--$10^4\,\tev)^{-2}$ (see the Appendix). Since the quark alignment
matrices $Q_{L,R}$ which diagonalize $\CM_q$ to $M_q$ are unimodular,
$\arg\det(M_q) = \arg\det(\CM_q) \equiv \arg\det(\CM_u) +
\arg\det(\CM_d)$. {\it Therefore, strong CP violation depends} entirely {\it
on the character of vacuum alignment in the technifermion sector---the
phases $\phi_{IJ}$ of $W$---and by how the ETC factors $\Lambda^{Tq}_{IijJ}$
map these phases into the $(\CM_q)_{ij}$.}

If the $\phi_{IJ}$ are random irrational phases, $\ol \theta_q$ could vanish
only by the most contrived, unnatural adjustment of the $\Lambda^{Tq}$. If
all $\phi_{IJ} = 2m\pi/N$ (mod $\pi$), then all elements of $\CM_u$ have the
same phase, as do all elements of $\CM_d$. Then, $U_{L,R}$ and $D_{L,R}$ will
be real orthogonal matrices, up to an overall phase. There may be strong
CP violation, but there will no weak CP violation in any interaction.

There remains the possibility, which we assume henceforth, that the
$\phi_{IJ}$ are different rational multiples of $\pi$. Then, strong CP
violation will be absent {\it IF} the $\Lambda^{Tq}$ map these phases onto
the primordial mass matrix so that (1) each element $(\CM_q)_{ij}$ has a
rational phase {\it AND} (2) these add to zero in $\arg\det(\CM_q)$. In the
absence of an explicit ETC model, we are not certain this can happen, but we
see no reason that it cannot. For example, there may be just one nonzero
$\Lambda^{Tq}_{IijJ}$ for each pair $(ij)$ and $(IJ)$. An ETC model which
achieves such a phase mapping will solve the strong CP problem, i.e., $\ol
\theta_q \simle 10^{-11}$, without an axion and without a massless up
quark. This is, in effect, a ``natural fine--tuning'' of phases in the quark
mass matrix: rational phase solutions are stable against substantial changes
in the nonzero $\Lambda^{TT}$. There is, of course, no reason weak CP
violation will not occur in this model. We shall illustrate this with some
examples in Sections~II.6 and~II.8.

Determining the quark alignment matrices $Q_{L,R}$ begins with minimizing the
vacuum energy
\be\label{eq:EqT}
E_{Tq}(Q) \cong -\half {\rm Tr}\left(\CM_q \ts Q + {\rm
    h.c.}\right)\Delta_q(\METC)
\ee
to find $Q=Q_L Q^\dag_R$. Whether or not $\ol \theta_q = 0$, the matrix
$Q^\dag \CM_q$ is hermitian up to the identity matrix~\cite{rfd},
\be\label{eq:nuyts}
 \CM_q \ts Q - Q^\dag \CM^\dag_q = i\nu_q \ts 1 \ts,
\ee
where $\nu_q$ is the Lagrange multiplier associated with the unimodularity
constraint on $Q$, and $\nu_q$ vanishes if $\ol \theta_q$ does. Therefore,
$\CM_q \ts Q$ may be diagonalized by the single unitary transformation $Q_R$
and, so,\footnote{Since quark vacuum alignment is based on first order chiral
perturbation theory, it is inapplicable to the heavy quarks $c,b,t$. When
$\ol \theta_q = 0$, Dashen's procedure is equivalent to making the mass
matrix diagonal, real, and positive. It then correctly determines the quark
unitary matrices $U_{L,R}$ and $D_{L,R}$ and the magnitude of strong and weak
CP violation.}
\be\label{eq:Mdiag}
M_q \equiv \left(\ba{cc} M_u & 0 \\ 0 & M_d \ea\right) = Q^\dag_R \CM_q
\ts Q Q_R = Q^\dag_R \CM_q \ts Q_L \ts.
\ee

\section*{II.6 Quark Mass and Mixing Matrices in ETC/TC2}

\subsection*{II.6.1 General Considerations}

From the block--diagonal $SU(6)$ matrices $Q_{L,R}$, one constructs the CKM
matrix $V= U^\dag_L D_L$. Carrying out the vectorial phase changes on the
$q_{L,R \ts i}$ required to put $V$ in the standard Harari--Leurer form with
the single CP--violating phase $\delta_{13}$, one obtains~\cite{harari,pdg}
\bea\label{eq:CKMmat}
V  &\equiv& \left(\ba{ccc}
      V_{ud} & V_{us} & V_{ub}\\
      V_{cd} & V_{cs} & V_{cb}\\
      V_{td} & V_{ts} & V_{tb}\\
      \ea\right)\\ \nn\\
 &=&  \left(\ba{lll}
      c_{12\ts} c_{13} 
      & s_{12\ts} c_{13}
      & s_{13\ts}e^{-i\delta_{13}}\\ 
      -s_{12\ts}c_{23}-c_{12\ts}s_{23\ts}s_{13\ts} e^{i\delta_{13}}
      & c_{12\ts}c_{23}-s_{12\ts}s_{23\ts}s_{13\ts} e^{i\delta_{13}}
      & s_{23\ts}c_{13\ts} \\
      s_{12\ts}s_{23}-c_{12\ts}c_{23\ts}s_{13\ts} e^{i\delta_{13}}
      & -c_{12\ts}s_{23}-s_{12\ts}c_{23\ts}s_{13\ts} e^{i\delta_{13}}
      & c_{23\ts}c_{13\ts}\\
      \ea\right) \ts.\nn
\eea
Here, $s_{ij} = \sin\theta_{ij}$, and the angles $\theta_{12}$,
$\theta_{23}$, $\theta_{13}$ lie in the first quadrant. Additional
CP--violating phases appear in $U_{L,R}$ and $D_{L,R}$ and they are rendered
observable  by ETC and TC2 interactions. We will study their contribution to
$\epsilon$ in Section~II.8. Before that, we need to discuss the
constraints on $\CM_{u,d}$ and on $U_{L,R}$ and $D_{L,R}$ imposed by ETC and
TC2.

First, as discussed in Section~I.3.1, limits on flavor--changing neutral
current (FCNC) interactions, especially those mediating $|\Delta S| =2$,
require that ETC bosons coupling to the two light generations have masses
$\METC \simge 1000\,\tev$. These can produce quark masses less than about
$m_s(\METC) \simeq 100\,\mev$ in a walking technicolor theory (see the
Appendix). Extended technicolor bosons as light as 50--$100\,\tev$ are needed
to generate $m_b(\METC) \simeq 3.5\,\gev$. Flavor--changing neutral current
interactions mediated by such light ETC bosons must be suppressed by small
mixing angles between the third and the first two generations.

The most important feature of $\CM_u$ is that the TC2 component of
$\CM_{tt}$, $(m_t)_{TC2} \simeq 160\,\gev$, is much larger than all its other
elements, all of which are generated by ETC exchange. In particular,
off-diagonal elements in the third row and column of $\CM_u$ are expected to
be no larger than the 0.01--1.0~GeV associated with $m_u$ and $m_c$. So,
$\CM_u$ is very nearly block--diagonal and $|U_{L,R \ts t u_i}| \cong
|U_{L,R \ts u_i t}| \cong \delta_{t u_i}$.

The matrix $\CM_d$ has a triangular or nearly triangular structure. One
reason for this is the need to suppress $\ol B_d$--$B_d$ mixing induced by
the exchange of ``bottom pions'' of mass $M_{\pi_b} \sim
300\,\gev$~\cite{kominis,bbhk}. Furthermore, since $U_L$ is block--diagonal,
the observed CKM mixing between the first two generations and the third must
come from the down sector. These requirements are met when the $d_R,s_R
\leftrightarrow b_L$ elements of $\CM_d$ are much smaller than the $d_L,s_L
\leftrightarrow b_R$ elements. In Ref.~\cite{tctwoklee}, the strong topcolor
$U(1)$ charges were chosen to exclude ETC interactions that induce $\CM_{db}$
and $\CM_{sb}$. This makes $D_R$, like $U_{L,R}$, nearly $2\times 2$ times
$1\times 1$ block--diagonal.

From these considerations and $V_{tb} \cong 1$, we have
\be\label{eq:ckm}
 V_{td_i} \cong V^*_{tb} \ts V_{td_i} \cong U_{L tt} D^*_{L bb}
  U^*_{L tt}  D_{L bd_i} \cong D^*_{L bb} D_{L bd_i} \ts.
\ee
This relation, which is good to 10\% (see Section~II.6.2 for examples), was
used in Ref.~\cite{blt} to put strong limits on the TC2 $V_8$ and $Z'$ masses
from $\ol B_d$--$B_d$ mixing (see Section~II.7). We found that $M_{V_8}$,
$M_{Z'} \simge 5\,\tev \gg (m_t)_{TC2}$. This implies that the TC2 gauge
couplings must be tuned to within 1\% or better of their critical values.
This stands as one the great challenges to TC2.

One more interesting property of the quark alignment matrices is this: The
vacuum energy $E_{Tq}$ is minimized when the elements of $U$ and $D$ have
almost the same rational phases as $\CM_u$ and $\CM_d$ do. In particular, all
the large diagonal elements of $U, D$ have rational phases (see
Section~II.6.2). This is generally not true of $U_{L,R}$ and $D_{L,R}$
individually. However, since $Q_{ii} = \sum_j Q_{L ij} Q^*_{R ij}$ ($Q =
U,D$) has a rational phase, $E_{Tq}$ is likely to be minimized when each term
in the sum has the same rational phase. Like DNA, in which the patterns
of the two strands are linked,
\be\label{eq:dna}
\arg Q_{L ij} - \arg Q_{L ik} = \arg Q_{R ij} - \arg Q_{R ik}
\quad ({\rm mod} \ts\ts \pi) 
\ee
for $i,j,k = u,c,t$ or $d,s,b$. In particular, $\arg V_{td_i} \cong \arg
D_{L bd_i} - \arg D_{L bb} \cong \arg D_{R bd_i} - \arg D_{R bb}$ (mod $\pi$)
for $d_i = d,s,b$.

\subsection*{II.6.2 Examples}

Our proposal for solving the strong CP problem in technicolor theories rests
on the fact that phases in the technifermion alignment matrices $W =
(W_U,W_D)$ can be different rational multiples of $\pi$, and on the
conjecture that these phases may be mapped by ETC onto the primordial mass
matrix $(\CM_q)_{ij} = \Lambda^{Tq}_{IijJ} \ts W^\dag_{JI} \Delta_T$ so that
$\ol \theta_q = \arg\det(\CM_q) = 0$. Corrections to $\ol \theta_q$ are then
expected to be at most $\CO(10^{-11})$. In this section we present two
examples of quark mass matrices for which we have engineered $\ol\theta_q
=0$. They lead to similar alignment and CKM matrices, except that one example
has $\delta_{13} =0$. Nevertheless, as we see in Section~II.8, both examples
lead to successful calculations of CP--violating parameter $\epsilon$.

\bigskip

\nin {\underbar{\it Model 1:}}

In this model, $\delta_{13} =0$, but CP violation will arise from phases in
$U_{L,R}$ and $D_{L,R}$. The primordial quark mass matrices renormalized at
$\METC$ are taken to be of seesaw form with phases that are multiples of
$\pi/3$:
\bea\label{eq:CMmodela}
\CM_u &=&\left(\ba{lll}        (0,\ts 0) & (200.,\ts 1/3)& (0,\ts 0)\\
                               (15.6,\ts -1/3) & (900,\ts 1) & (0,\ts 0)\\
                               (0,\ts 0) & (0,\ts 0) & (162620,\ts 0)\ea\right)
                               \nn\\ \\
\CM_d &=&\left(\ba{lll}        (0,\ts 0) & (23.3,\ts 0) & (0,\ts 0)\\
                               (21.7,\ts 0) & (102,\ts 1/3) & (0,\ts 0)\\
                               (17.0,\ts 1/3) & (144,\ts 2/3) & (3505,\ts
                               0)\ea\right)
                               \nn \ts.
\eea
The notation is $(|(\CM_q)_{ij}|, \ts \arg [(\CM_q)_{ij}]/\pi)$. The mass
units are MeV. Here we made $\arg\det(\CM_u) = \arg\det(\CM_d) = \pi$. We
imposed the same kind of structure on $\CM_u$ as $\ol B_d$--$B_d$ mixing
requires of $\CM_d$. The quark mass eigenvalues may be extracted from
$\CM_q$. Their values at $\METC \sim 10^3\,\tev$ are:
\bea\label{eq:mqmodela}
m_u &=& 3.35\ts, \ts\ts\ts m_c = 924\ts, \ts\ts\ts m_t = 162620 \nn\\
m_d &=& 4.74\ts, \ts\ts\ts m_s = 106\ts, \ts\ts\ts m_b = 3508
\eea

The alignment matrices $U = U^\dag_L U_R$ and $D = D^\dag_L D_R$ obtained by
minimizing $E_{Tq}$ are
\bea\label{eq:Qmodela}
U &=&\left(\ba{lll}     (0.973,\ts 0) & (0.232,\ts 1/3) & (0,\ts 0)\\
                        (0.232,\ts -1/3) & (0.973,\ts 1)& (0,\ts 0)\\
                        (0,\ts 0) & (0,\ts 0) & (1,\ts 0)\ea\right)
                        \nn\\ \\
D &=&\left(\ba{lll}     (0.915,\ts -2/3) & (0.404,\ts 0) & (0.0046,\ts -1/3)\\
                        (0.404,\ts 0) & (0.914,\ts -1/3)& (0.0400,\ts -2/3)\\
                        (0.0119,\ts -1/3) & (0.0384,\ts -2/3) & (0.999,\ts
                        0)\ea\right)
                        \nn \ts.
\eea
The cloning of the $\CM_{u,d}$ phases onto $U,D$ is apparent. Diagonalizing
the aligned quark mass matrices yields $Q_{L,R}$:
\bea\label{eq:QLRmodela}
U_L &=&\left(\ba{lll}   (0.9999,\ts -0.859) & (0.0164,\ts 0.141) & (0,\ts 0)\\
                        (0.164,\ts -1.193) & (0.9999,\ts -1.193) & (0,\ts 0)\\
                        (0,\ts 0) & (0,\ts 0) & (1,\ts -0.526)\ea\right) 
                        \nn\\ \nn \\
D_L &=&\left(\ba{lll}   (0.980,\ts 1.141) & (0.199,\ts 1.141) & (0.00485,\ts
                          1.141)\\ 
                        (0.199,\ts -0.192) & (0.979,\ts 0.808) & (0.0412,\ts
                        0.808)\\ 
                        (0.00344,\ts -0.526) & (0.0413,\ts 0.474) &
                        (0.999,\ts -0.526) \ea\right)
                        \nn\\ \\
U_R &=&\left(\ba{lll}   (0.976,\ts -0.859) & (0.216,\ts -0.859) & (0,\ts 0)\\
                        (0.216,\ts -1.193) & (0.976,\ts -0.192) & (0,\ts 0)\\
                        (0,\ts 0) & (0,\ts 0) & (1,\ts -0.526)\ea\right)
                        \nn\\ \nn\\ 
D_R &=&\left(\ba{lll}   (0.977,\ts -0.192) & (0.214,\ts 0.808) &(0.000273,\ts
                          0.808)\\
                        (0.214,\ts 1.141) & (0.977,\ts 1.141) &(0.00122,\ts
                        1.141)\\  
                     (5\times 10^{-6},\ts -0.526) & (0.00125,\ts 0.474)
                     &(1,\ts -0.526) \ea\right) \ts.\nn
\eea
As required, all the mixing in $U_{L,R}$ and $D_R$ is between the first two
generations; mixing of these two with the third generation comes entirely
from $D_L$. A perusal of the phases will reveal differences which are
multiples of $\pi/3$. Finally, the CKM matrix is
\be\label{eq:CKMmodela}
V = \left(\ba{lll}   (0.977,\ts 0) & (0.215,\ts 0) & (0.00552,\ts 0)\\
                        (0.215,\ts 1) & (0.976,\ts 0) & (0.0411,\ts 0)\\
                        (0.00344,\ts 0) & (0.0413,\ts 1) & (0.999,\ts 0)
                        \ea\right) \ts.
\ee
Note its similarity to $D_L$ (including phase differences). This corresponds
to the angles
\be\label{eq:CKMangsmodela}
\theta_{12} = 0.217 \ts, \ts\ts\ts \theta_{23} = 0.0411 \ts, \ts\ts\ts
\theta_{13} = 0.00552 \ts, \ts\ts\ts \delta_{13} = 0\ts.
\ee
The angles $\theta_{ij}$ are in good agreement with those in the Particle
Data Group's book~\cite{pdg}. We will see in Section~II.8 that, even though
$\delta_{13} = 0$, the CP--violating angles in $D_{L,R}$ can easily account
for the measured value of $\epsilon$.

\bigskip

\nin {\underbar{\it Model 2:}}

The second model is based on a $W$--matrix whose phases are multiples of
$\pi/5$. The primordial quark mass matrices renormalized at $\METC$ are again
taken to be of seesaw form, but we allow off--diagonal terms $|\CM_{ij}| \sim
\sqrt{(|\CM_{ii} \CM_{jj}|)}$ (all masses refer to the ETC contribution only):
\bea\label{eq:CMmodelb}
\CM_u &=&\left(\ba{lll}        (7,\ts 0.2) & (2,\ts -0.4)& (0,\ts 0)\\
                               (100,\ts 0.4) & (890,\ts -0.2) & (0,\ts 0)\\
                               (50,\ts -0.4) & (500,\ts 0.2) & (160000,\ts
                               0)\ea\right)
                               \nn\\ \\
\CM_d &=&\left(\ba{lll}        (8,\ts 0) & (1,\ts -0.2) & (0,\ts 0)\\
                               (25,\ts -0.2) & (100,\ts -0.4) & (0,\ts 0)\\
                               (10,\ts 0) & (140,\ts -0.4) & (3500,\ts
                               0.4)\ea\right)
                               \nn \ts.
\eea
Here we made $\arg\det(\CM_u) = \arg\det(\CM_d) = 0$. We again imposed
the same kind of structure on $\CM_u$ as $\ol B_d$--$B_d$ mixing requires of
$\CM_d$. The quark mass eigenvalues are:
\bea\label{eq:mqmodelb}
m_u &=& 6.84\ts, \ts\ts\ts m_c = 896\ts, \ts\ts\ts m_t = 160000\nn\\
m_d &=& 7.52\ts, \ts\ts\ts m_s = 103\ts, \ts\ts\ts m_b = 3503
\eea

The alignment matrices $U = U^\dag_L U_R$ and $D = D^\dag_L D_R$ obtained by
minimizing $E_{Tq}$ are
\bea\label{eq:Qmodelb}
U &=&\left(\ba{lll}     (0.994,\ts -0.2) & (0.110,\ts -0.4) & (0.00031,\ts
                          0.4)\\ 
                        (0.110,\ts -0.6) & (0.994,\ts 0.2)& (0.00311,\ts
                        -0.2)\\ 
                        (0.00062,\ts 0.505) & (0.00306,\ts -0.6) & (1,\ts
                        0)\ea\right) 
                        \nn\\ \\
D &=&\left(\ba{lll}     (0.976,\ts 0) & (0.217,\ts 0.2) & (0.00265,\ts
                         -0.0178)\\ 
                        (0.217,\ts -0.8) & (0.975,\ts 0.4)& (0.0389,\ts
                        0.4)\\ 
                        (0.00664,\ts -0.679) & (0.0384,\ts 0.603) &
                        (0.999,\ts -0.4)\ea\right)
                        \nn \ts.
\eea
Again, the cloning of the $\CM_{u,d}$ phases onto the large elements of $U,D$
is apparent. The $Q_{L,R}$ are:
\bea\label{eq:QLRmodelb}
U_L &=&\left(\ba{lll}   (0.994,\ts 0.873) & (0.112,\ts 0.336) & (0.00031,\ts
                          0.535)\\ 
                        (0.112,\ts 0.472) & (0.994,\ts 0.936) & (0.00313,\ts
                        -0.0652)\\ 
                        (0.00063,\ts -0.422) & (0.00308,\ts 0.138) &
                        (1,\ts 0.135)\ea\right)
                        \nn\\ \nn \\
D_L &=&\left(\ba{lll}   (0.970,\ts 0.881) & (0.245,\ts 0.727) & (0.00286,\ts
                          0.535)\\ 
                        (0.245,\ts 0.0810) & (0.969,\ts 0.927) & (0.0400,\ts
                        0.936)\\ 
                        (0.00771,\ts 0.213) & (0.0394,\ts 1.131) & (0.999,\ts
                        0.135) \ea\right)
                        \nn\\ \\
U_R &=&\left(\ba{lll}   (1,\ts 1.073) & (0.00198,\ts 0.534) & (0,\ts 0)\\
                        (0.00198,\ts 0.274) & (1,\ts 0.736) & (1.8\times
                        10^{-5},\ts -0.322)\\ 
                        (0,\ts 0) & (1.8\times 10^{-5},\ts 0.192)& (1,\ts
                        0.135)\ea\right)
                        \nn\\ \nn\\ 
D_R &=&\left(\ba{lll}   (1,\ts 0.881) & (0.0284,\ts 0.727) &(1.7\times
                         10^{-5},\ts 0.661)\\ 
                        (0.0284,\ts -0.319) & (1,\ts 0.527) &(0.00116,\ts
                         0.531)\\ 
                     (1.7\times 10^{-5},\ts 0.611) & (0.00116,\ts -0.470)
                         &(1,\ts 0.535) 
                        \ea\right) \ts.\nn
\eea
The CKM matrix is (compare it to $D_L$)
\be\label{eq:CKMmodelb}
V = \left(\ba{lll}   (0.972,\ts 0) & (0.234,\ts 0) & (0.00315,\ts 0.305)\\
                     (0.233,\ts 0.9999) & (0.971,\ts 8.6\times 10^{-6}) &
                     (0.0431,\ts 0)\\ 
                     (0.00867,\ts 0.0930) & (0.0423,\ts 0.995) & (0.999,\ts 0)
                        \ea\right) \ts.
\ee
This corresponds to the angles
\be\label{eq:CKMangsmodelb}
\theta_{12} = 0.236 \ts, \ts\ts\ts \theta_{23} = 0.0431 \ts, \ts\ts\ts
\theta_{13} = 0.00315 \ts, \ts\ts\ts \delta_{13} = -0.957\ts.
\ee
Again, the angles $\theta_{ij}$ are in reasonable agreement with those
in the Particle Data Group's book. In this model, $\delta_{13}$ is large.

\section*{II.7 ETC and TC2 Four--Fermion Interactions} 

The FCNC effects that concern us arise from four--quark interactions induced
by the exchange of heavy ETC gauge bosons and of TC2 color--octet ``colorons''
$V_8$ and color--singlet $Z'$. Lepton interactions are ignored.

At low energies and to lowest order in $\alpha_{ETC}$, the ETC interaction
involves products of chiral currents. Still assuming that the ETC gauge
group commutes with electroweak $SU(2)$, it has the form
\bea\label{eq:HETC}
\chetc &=& \Lambda^{LL}_{ijkl} \left(\ol u'_{Li} \gamma^\mu u'_{Lj} + \ol
d^{\ts \prime}_{Li} \gamma^\mu d^{\ts \prime}_{Lj}\right) \left(\ol u'_{Lk}
\gamma^\mu u'_{Ll} + \ol d^{\ts \prime}_{Lk} \gamma^\mu d^{\ts
\prime}_{Ll}\right) \nn\\
& & \ts +
\left(\ol u'_{Li} \gamma^\mu u'_{Lj} + \ol d^{\ts \prime}_{Li} \gamma^\mu 
d^{\ts \prime}_{Lj}\right)
\left(\Lambda^{u,LR}_{ijkl} \ol u'_{Rk}\gamma^\mu u'_{Rl} +
      \Lambda^{d,LR}_{ijkl} \ol d^{\ts \prime}_{Rk}\gamma^\mu
      d^{\ts \prime}_{Rl}\right) \nn \\
& & \ts + \Lambda^{uu,RR}_{ijkl}\ol u'_{Ri}\gamma^\mu u'_{Rj}\ts\ol
u'_{Rk}\gamma^\mu u'_{Rl} + \Lambda^{dd,RR}_{ijkl}\ol d^{\ts
\prime}_{Ri}\gamma^\mu d^{\ts \prime}_{Rj} \ts\ol d^{\ts
\prime}_{Rk}\gamma^\mu d^{\ts \prime}_{Rl} \nn \\
& & \ts + \Lambda^{ud,RR}_{ijkl} \ol u'_{Ri}\gamma^\mu u'_{Rj}\ts
\ol d^{\ts \prime}_{Rk}\gamma^\mu d^{\ts \prime}_{Rl} \ts,
\eea
where primed fields are electroweak eigenstates. The ETC gauge group contains
technicolor, color and topcolor, and flavor as commuting
subgroups~\cite{etceekl}. It follows that the flavor currents in $\chetc$ are
color and topcolor singlets. The $\Lambda$'s in $\chetc$ are of order
$\getc^2/\METC^2$, whose magnitude is discussed below, and the operators are
renormalized at $\METC$. Hermiticity of $\chetc$ implies that $\Lambda_{ijkl}
= \Lambda^*_{jilk}$. We assume that this primordial ETC interaction conserves
CP, i.e., that all the $\Lambda$'s are real. When written in terms of mass
eigenstate fields $q_{L,R\ts i} = \sum_j (Q^\dag_{L,R})_{ij} q'_{L,R\ts j}$
with $Q = U,D$, an individual four--quark term in $\chetc$ has the form
\be\label{Hqterm}
  \left(\sum_{i'j'k'l'} \Lambda^{q_1 q_2 \lambda_1 \lambda_2}_{i'j'k'l'}
  \ts\ts   Q^\dag_{\lambda_1\ts ii'} Q_{\lambda_1\ts j'j} \ts
  Q^\dag_{\lambda_2\ts kk'} Q_{\lambda_2\ts l'l}\right) \ts 
  \ol q_{\lambda_1 i} \ts \gamma^\mu \ts q_{\lambda_1 j} \ts\ts
  \ol q_{\lambda_2 k} \ts \gamma_\mu \ts q_{\lambda_2 l} \ts. 
\ee

A reasonable and time--honored guess for the magnitude of the
$\Lambda_{ijkl}$ is that they are comparable to the ETC masses that generate
the quark mass matrix $\CM_q$. We elevate this to a rule: The ETC scale
$\METC/\getc$ in a term involving weak eigenstates of the form $\ol q^{\ts
\prime}_i q'_j \ol q^{\ts \prime}_j q'_i$ or $\ol q^{\ts \prime}_i q'_i \ol
q^{\ts \prime}_j q'_j$ (for $q'_i = u'_i$ or $d^{\ts \prime}_i$) is
approximately the same as the scale that generates the $\ol q^{\ts
\prime}_{Ri} q'_{Lj}$ mass term, $(\CM_q)_{ij}$. A plausible, but
approximate, scheme for correlating a quark mass $m_q(\METC)$ with
$\METC/\getc$ is presented in the Appendix. The results are shown in
Figure~2.  There, $\kappa > 1$ parameterizes the departure from the strict
walking technicolor limit. That is, we take $\atc =$ constant and the
anomalous dimension $\gamma_m$ of $\ol T T$ equal one up to the highest ETC
mass scale (the one generating $m_{u,d} \sim$ few MeV) divided by $\kappa^2$,
and $\gamma_m = 0$ beyond that. The ETC masses run from $\METC/\getc =
46\,\tev$ for $m_q = 5\,\gev$ to $2.33/\kappa\times 10^4\,\tev$ for $m_q =
10\,\mev$. We rely on Figure~2 for estimating the $\Lambda$'s in $\chetc$, in
particular, for the calculations of $\epsilon$ in Section~II.8.

\begin{figure}[t]
 \vspace{6.0cm}
\includegraphics{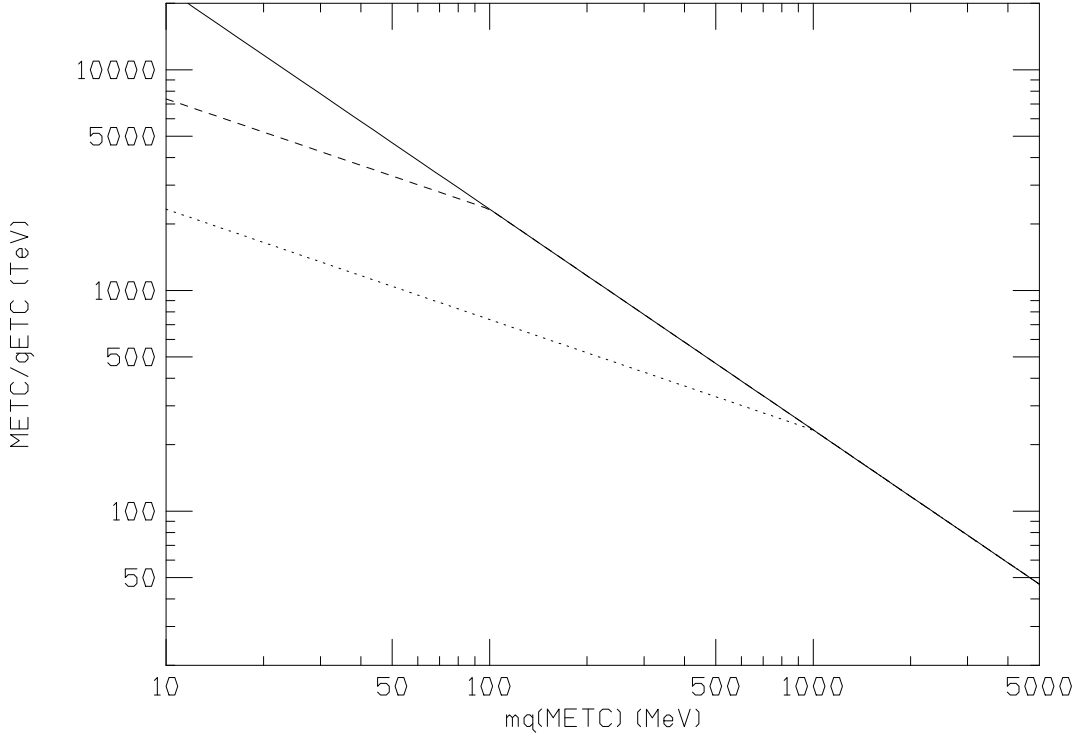}
\vskip 4.0truecm
 \caption{\it
   Extended technicolor scale $\METC/\getc$ as a function of quark mass
   $m_q$ renormalized at $\METC$ for $\kappa = 1$ (solid curve), $\sqrt{10}$
   (dashed), and $10$ (solid); see the Appendix for details.
    \label{fig2} }
\end{figure}

Very large extended technicolor masses are necessary, but not sufficient, to
suppress FCNC interactions of light quarks to an acceptable level. This is
especially true for $\Delta M{_K^0}$ and $\epsilon$. We must also assume that
$\chetc$ is {\it electroweak generation conserving}, i.e.,
\be\label{eq:gencon}
\Lambda^{q_1q_2\lambda_1\lambda_2}_{ijkl} = \delta_{il}\delta_{jk}
\Lambda^{q_1q_2\lambda_1\lambda_2}_{ij} +  \delta_{ij}\delta_{kl}
\Lambda^{\prime \ts q_1q_2\lambda_1\lambda_2}_{ik}\ts.
\ee
Considerable FCNC suppression then comes from off--diagonal elements in the
alignment matrices $Q_{L,R}$.

In all TC2 models, color $SU(3)_C$ and weak hypercharge $\uy$ arise from the
breakdown of the topcolor groups $\suone\otimes\sutwo$ and
$\uone\otimes\utwo$ to their diagonal subgroups. Here, $\suone$ and $\uone$
are strongly coupled, $\sutwo$ and  $\utwo$ are weakly coupled, with the
color and weak hypercharge couplings given by
\bea\label{eq:tccouplings}
g_C &=& {g_1 g_2 \over{\sqrt{g_1^2 + g_2^2}}} \equiv {g_1 g_2\over{g_{V_8}}}
    \equiv g_2 \cos\thc \simeq g_2 \ts; \quad \nn\\
g_Y &=& {g'_1 g'_2 \over{\sqrt{g_1^{\prime \ts 2} + g_2^{\prime \ts 2}}}}
\equiv {g'_1 g'_2\over{g_{Y}}} \equiv g'_2 \cos\thy \simeq g'_2 \ts.
\eea
Top and bottom quarks are $\suone$ triplets. The broken topcolor interactions
are mediated by a color octet of colorons, $V_8$, and a color singlet $Z'$
boson, respectively. By virtue of the different $\uone$ couplings of $t_R$
and $b_R$, exchange of $V_8$ and $Z'$ between third generation quarks
generates a large contribution, $(m_t)_{TC2} \simeq 160\,\gev$, to the
top mass, but none to the bottom mass.

If topcolor is to provide a natural explanation of $(m_t)_{TC2}$, the $V_8$
and $Z'$ masses ought to be $\CO(1\,\tev)$. In the Nambu--Jona-Lasinio (NJL)
approximation for the $V_8$ and $Z'$--exchange interactions, the degree to
which this naturalness criterion is met is quantified by the ratio~\cite{cdt}
\be\label{eq:tune}
{\alpha(V_8) + \alpha(Z') - (\alpha^*(V_8) + \alpha^*(Z'))\over{
\alpha^*(V_8) + \alpha^*(Z')}} = {\alpha(V_8)\ts r_{V_8} + \alpha(Z')\ts
r_{Z'} \over {\alpha(V_8)(1-r_{V_8}) + \alpha(Z')(1-r_{Z'})}} \ts.
\ee
Here,
\bea\label{eq:rdef}
\alpha(V_8) &=& {4\alpha_{V_8} \cos^4\thc\over{3\pi}}
         \equiv {4\alpha_C \cot^2\thc\over{3\pi}}, \nn\\
\alpha(Z')  &=& {\alpha_{Z'} Y_{t_L} Y_{t_R} \cos^4\thy\over{\pi}} 
         \equiv {\alpha_Y Y_{t_L} Y_{t_R} \cot^2\thy\over{\pi}}\ts;\\
\tan\thc &=& {g_2\over{g_1}} \ts, \quad \tan\thy = {g'_2\over{g'_1}} \ts,
\quad
r_i = {(m^2_t)_{TC2}\over{M^2_i}} \ts \ln \left({M^2_i
\over{(m^2_t)_{TC2}}}\right) \ts, \hskip0.25truein (i = V_8, Z')\ts;\nn
\eea
and $Y_{t_{L,R}}$ are the $\uone$ charges of $t_{L,R}$. The NJL condition on
the critical couplings for top condensation is $\alpha^*(V_8) + \alpha^*(Z')
= 1$. In Ref.~\cite{blt} we showed that, for such large couplings, TC2 is
tightly constrained by the magnitude of $\ol B_d$--$B_d$ mixing, requiring
$M_{V_8} \simeq M_Z' \simge 5\,\tev$. This implies that the topcolor coupling
$\alpha(V_8) + \alpha(Z')$ must be within less than 1\% of its critical
value, a tuning we regard as unnaturally fine. Other limits on $M_{V_8}$ were
obtained in Refs.~\cite{others,bk}. It may be possible to eliminate this
fine--tuning problem by using flavor--universal TC2~\cite{ccs,ehsbbmix}.
Another possibility is to invoke the ``top seesaw'' mechanism in which the
topcolor interactions operate on a quark whose mass is several TeV, and the
top's mass comes to it by a seesaw mechanism~\cite{seesaw}.

In standard TC2~\cite{tctwohill}, $V_8$ and $Z'$ exchange also give rise to
FCNC that mediate $|\Delta S| =2$ and $|\Delta B|= 2$ processes. In
flavor--universal TC2, only $Z'$ exchange generates such FCNC. We shall write
the four--quark interaction for standard TC2, but our results apply to $Z'$
exchange interactions in flavor--universal TC2 as well. How large the FCNC
rates are there depends on the strength of the $Z'$
couplings~\cite{ehsbbmix}.

The TC2 interaction at energies well below $\Mv$ and $\Mzp$ is
\be\label{eq:HTCT}
\chtct = {g^2_{V_8} \over{2 M^2_{V_8}}} \sum_{A=1}^8 J^{A\mu} J^A_\mu + 
 {g^2_{Z'} \over{2 M^2_{Z'}}} J_{Z'}^\mu J_{Z'\mu} \ts.
\ee
The coloron and $Z'$ currents written in terms of electroweak eigenstate
fields are given by (color indices are suppressed)
\bea\label{eq:JTCT}
 J^A_\mu &=& \cos^2\thc \sum_{i=t,b} \ol q'_i \gamma_\mu \ts
 t_A \ts  q'_i - \sin^2\thc \sum_{i=u,d,c,s} \ol q'_i
 \gamma_\mu\ts  t_A \ts q'_i \ts; \nn\\
 J_{Z'\mu} &=& \cos^2\thy J_{1\mu} - \sin^2\thy J_{2\mu} \\
     &\equiv& \sum_{\lambda =L,R} \sum_i \left(\cos^2\thy \ts Y_{1\lambda i}
             - \sin^2\thy \ts Y_{2\lambda i} \right)
           \ol q'_{\lambda i} \gamma_\mu q'_{\lambda i} \ts\ts.\nn
\eea
The $\uone$ and $\utwo$ hypercharges satisfy $Y_{1\lambda i} + Y_{2\lambda i}
= Y_{\lambda i} = 1/6$ or $Q_{EM}$ for $\lambda = L$ or $R$. Consistency with
$SU(2)$ symmetry requires $Y_{Lt} = Y_{Lb}$, etc. The suppression of light
quark FCNC requires $Y_{1Li} \equiv Y_{1i}$ for $i=u,d,c,s$ and $Y_{1Ru} =
Y_{1Rc}$, $Y_{1Rd} = Y_{1Rs}$. Remaining FCNC will have to be--and
are--suppressed by small mixing angles between the first two generations and
the third.

\section*{II.8 Contributions to Weak CP Violation}

The CP--violating parameter $\epsilon$ in the $K^0$ system is defined by
\be\label{eq:epsdef}
\epsilon \equiv {A(K_L \ra (\pi\pi)_{I=0}) \over {A(K_S \ra (\pi\pi)_{I=0})}}
= {e^{i\pi/4} \ts {\rm Im} M_{12} \over{\sqrt{2}\ts \Delta M_K}},
\ee
where $2 M_K M_{12} = \langle K^0| \CH_{|\Delta S| = 2} |\ol K^0\rangle$ and
we use the phase convention that $A_0 = \langle (\pi\pi)_{I=0}| \CH_{|\Delta
S| = 1} |K^0\rangle$ is real. Experimentally~\cite{pdg},
\be\label{eq:epsvalue}
\epsilon = (2.271\pm 0.017)\times 10^{-3} \exp{(i\pi/4)}\ts.
\ee
The standard model contribution to $\epsilon$ is~\cite{buras}
\bea\label{eq:epssm}
\epsilon_{SM} &=& {e^{i\pi/4} \ts G_F^2 M_W^2 f_K^2 \hat B_K M_K
  \over{3\sqrt{2}\pi^2 \Delta M_K}}\\
&\times&{\rm Im}\left[\lambda_c^{*\ts 2} \eta_1 S_0(x_c)
             + \lambda_t^{*\ts 2} \eta_2 S_0(x_t)
             + 2\lambda_c^*\lambda_t^* \eta_3 S_0(x_c,x_t)\right] \ts,\nn
\eea
where $f_K = 112\,\mev$ is the kaon decay constant, $\hat B_K = 0.80 \pm
0.15$ is the kaon bag parameter, $\lambda_{i=c,t} = V_{id} V^*_{is}$, and the
other quantities are defined in Ref.~\cite{buras}.

Despite the large ETC gauge boson masses of several 1000~TeV and the
stringent $\ol B_d$--$B_d$ mixing constraint leading to TC2 gauge masses of at
least 5~TeV, both interactions can contribute significantly to
$\epsilon$. The main ETC contribution comes from $\ol s' s' \ol s' s'$ 
interactions and is given by 
\bea\label{eq:epsetc}
\epsilon_{ETC} &\simeq& {e^{i\pi/4} \ts f_K^2 M_K \hat B_K\over{3\sqrt{2}\ts
    \Delta M_K}} \nn\\
&\times&  \Biggl\{-\left[\left({M_K\over{m_s + m_d}}\right)^2 +
  {3\over{2}}\right] 
      \Lambda^{LR}_{ss} \ts {\rm Im}\left(D_{Lss} D^*_{Lsd} D_{Rss} D^*_{Rsd}
      \right) \nn\\
&&\hskip0.1in + \ts 2\left[\Lambda^{LL}_{ss} \ts {\rm Im}\left(D^2_{Lss}
    D^{*\ts 2}_{Lsd} 
           \right)
           +\Lambda^{RR}_{ss} \ts {\rm Im}\left(D^2_{Rss} D^{*\ts
               2}_{Rsd}\right)\right]\Biggr\} \ts.
\eea
Note the suppression of $\CO((\theta_{12})^2)$ from mixing angle
factors. This $\ol s' s' \ol s' s'$ contribution, as well as those from the
standard model and TC2, vanish for Model 1. For that model, ${\rm
Im}(M_{12})_{ETC}$ comes from $\ol s' d' \ol d' s'$ ETC terms and has a form
similar to Eq.~(\ref{eq:epsetc}).

The dominant (standard) TC2 contribution comes from $\ol b'_L b'_L \ol b'_L
b'_L$ interactions; terms involving $b'_R$ are suppressed by the very small
$D_{Rbd}$ and $D_{Rbs}$:
\bea\label{eq:epstct}
\epsilon_{TC2} &\simeq& {e^{i\pi/4} \ts 4 \pi f_K^2 M_K \hat
                       B_K\over{3\sqrt{2}\ts \Delta M_K}} \nn\\
&\times& \left[{\alpha_C \cot^2\thc \over{M^2_{V_8}}} + 
              {\alpha_Y (\Delta Y_L)^2 \cot^2\thy \over{M^2_{Z'}}}\right]
             \ts {\rm Im}\left(D^2_{Lbs} D^{*\ts 2}_{Lbd}\right)
             \ts.
\eea
The couplings and mixing angles were defined in Eq.~(\ref{eq:rdef}) and
$\Delta Y_L = Y_{b_L} - Y_{d_L} = Y_{b_L} - Y_{s_L}$ is a difference of
strong $\uone$ hypercharges. Finally, following Ref.~\cite{blt}, we take
$\alpha_C \cot^2\thc = \alpha_Y (\Delta Y_L)^2 \cot^2\thy = 3\pi/8$.

{\begin{table}[t]
\begin{center}
%
\begin{tabular}{|c|c|c|c|c|c|c|c|}
\hline
Model& SM & ${\rm (ETC)_{LR}}$ & ${\rm (ETC)_{LL}}$ & ${\rm (ETC)_{RR}}$ &
${\rm (TC2)_{LL}}$ & $\delta_{13}$ & $\sin(2\beta)$\\
\hline\hline
1 & 0 & 2.38 & 0 & 0 & 0 & 0 & 0\\
\hline
1' & 2.28 & 9.61 & 0.88 & 1.02 & 8.34 & 0.98 & 0.98\\
\hline
2 & -1.98 & 9.44 & -7.68 & -0.11 & -4.57 & -0.96 & -0.55\\
\hline
2' & 1.97 & -6.30 & 7.76 & 0.05 & 4.52 & 0.96 & 0.55\\
\hline
2'' & -2.02 & 31.25 & -7.92 & -1.21 & -4.68 & -0.96 & -0.55\\
\hline
3 & 2.18 & -8.94 & -0.97 & -0.80 & 8.20 & 0.98 & 0.86\\
\hline\hline
\end{tabular}
\caption{Contributions to $\epsilon \ts e^{-i\pi/4} \times
  10^3$, the CKM phase $\delta_{13}$, and $\sin(2\beta)$ for various
  ``models'' of $\CM_q$ with $\ol \theta_q = 0$. Unless noted  otherwise in
  the text, all $\Lambda_{ss} = (2000\,\tev)^{-2}$ and $M_{V_8} = M_{Z'} =
  10\,\tev$. \label{tab:epsil}}
\end{center}
%
\end{table}}

The various contributions to $\epsilon$ and the CKM phase $\delta_{13}$ for
several different ``models'' of the primordial quark mass matrix $\CM_q$ are
given in Table~1. In all cases, $\epsilon$--contributions were calculated
using $\Lambda_{ss} = (2000\,\tev)^{-2}$ and $M_{V_8} = M_{Z'} =
10\,\tev$. Models~1 and~2 discussed above are present as are some related
ones (e.g., model~2' is similar to model~2, but the complex conjugate input
$\CM_q$ is used.  Differences apart from signs between models~2 and~2' are
due to computer round--off error).

In model~1, we used $\Lambda_{sd} = (4000\,\tev)^{-2}$. We see that this
model accounts surprisingly well for $\epsilon$ from ETC interactions alone!
Unfortunately, it gives $\sin(2\beta) =0$ (discussed below) and too small a
value for $\epsilon'/\epsilon$. The standard model contribution to $\epsilon$
in models~1' and~2' saturate its experimental value. Any reasonable choice of
ETC and TC2 masses spoils this agreement, so that these ETC/TC2 models are
ruled out. In model~2, the choice $\Lambda_{ss} = (1250\,\tev)^{-2}$ and
$M_{V_8}, M_{Z'} \ra \infty$ give $\epsilon({\rm ETC} + {\rm TC2}) = 4.22$,
and $\epsilon = 2.24$ (always times $10^{-3} \exp{(i\pi/4)}$). In model~2'',
$\epsilon({\rm ETC} + {\rm TC2}) = 4.33$, so that $\epsilon = 2.29$, for
$\Lambda_{ss} = (2500\,\tev)^{-2}$ and $M_{V_8}=M_{Z'} = 6.9\,\tev$. In
model~3, $\epsilon({\rm ETC} + {\rm TC2}) = 0.10$ and $\epsilon = 2.28$ for
$\Lambda_{ss} = (2000\,\tev)^{-2}$ and $M_{V_8}=M_{Z'} = 8.7\,\tev$. In these
three models, a good value of $\epsilon$ is obtained because large
cancellations occur among the ETC contributions or between ETC and TC2
contributions. These cancellations might seem contrived, but the same thing
happens in the standard model---between the QCD and electroweak penguin terms
in the calculation of $\epsilon'/\epsilon$~\cite{buras}---and we have learned
to live with that.

Fairly precise measurements of CP violation in the $\ol B_d$--$B_d$ system
have been made out by the Babar and Belle Collaborations~\cite{CPBB}.
Assuming this CP violation comes entirely from the CKM matrix, with no ETC
and TC2 contributions, it is characterized by the angle $\beta$ given by
\be\label{eq:sintwobeta}
\tan \beta = {{\rm Im}V^*_{td}\over{{\rm Re}V^*_{td}}}\ts.
\ee
Babar and Belle find
\bea\label{eq:stwobmeas}
\sin(2\beta) &=&  0.59 \pm 0.14 \ts\ts {\rm(stat)}\ts\ts \pm 0.05 \ts\ts
{\rm(syst)} \quad {\rm (Babar)} \ts;\nn\\
             &=&  0.99 \pm 0.14 \ts\ts {\rm(stat)}\ts\ts \pm 0.06 \ts\ts
{\rm(syst)} \quad {\rm (Belle)} \ts.
\eea
Only model~3 accomodates these measurements and the value of $\epsilon$; they
are a powerful discriminator of the CP--violation mechanism we propose.

For the curious, here are the mass and mixing matrices for model~3. Note how
similar they are to model~1.

\nin {\underbar{\it Model 3:}}
\bea\label{eq:CMmodelc}
\CM_u &=&\left(\ba{lll}        (0,\ts 0) & (200.,\ts 0)& (0,\ts 0)\\
                               (16,\ts 2/3) & (900,\ts 0) & (0,\ts 0)\\
                               (0,\ts 0) & (0,\ts 0) & (160000,\ts 0)\ea\right)
                               \nn\\ \\
\CM_d &=&\left(\ba{lll}        (0,\ts 0) & (20,\ts -1/3) & (0,\ts 0)\\
                               (22,\ts 0) & (100,\ts 0) & (0,\ts 0)\\
                               (17.0,\ts 0) & (145,\ts -1/3) & (3500,\ts
                               -1/3)\ea\right)
                               \nn \ts.
\eea
This differs from model~1 only in the choice of phases. The quark mass
eigenvalues are:
\bea\label{eq:mqmodelc}
m_u &=& 3.47\ts, \ts\ts\ts m_c = 922\ts, \ts\ts\ts m_t = 160000 \nn\\
m_d &=& 4.22\ts, \ts\ts\ts m_s = 104\ts, \ts\ts\ts m_b = 3503
\eea

The alignment matrices $U = U^\dag_L U_R$ and $D = D^\dag_L D_R$ obtained by
minimizing $E_{Tq}$ are
\bea\label{eq:Qmodelc}
U &=&\left(\ba{lll}     (0.972,\ts 1/3) & (0.233,\ts -2/3) & (0,\ts 0)\\
                        (0.233,\ts 0) & (0.972,\ts 0)& (0,\ts 0)\\
                        (0,\ts 0) & (0,\ts 0) & (1,\ts 0)\ea\right)
                        \nn\\ \\
D &=&\left(\ba{lll}     (0.922,\ts -2/3) & (0.387 ,\ts 0) & (0.0047,\ts
                        -0.014)\\ 
                        (0.387,\ts 1/3) & (0.921,\ts 0)& (0.0402,\ts 1/3)\\
                        (0.0141,\ts -0.755) & (0.0379,\ts -0.987) & (0.999,\ts
                        1/3)\ea\right)
                        \nn \ts.
\eea
The cloning of the $\CM_{u,d}$ phases onto $U,D$ is again apparent, but
phases of terms with small magnitudes are not well determined
numerically. Diagonalizing the aligned quark mass matrices yields $Q_{L,R}$:
\bea\label{eq:QLRmodelc}
U_L &=&\left(\ba{lll}   (0.9999,\ts 0.093) & (0.0169,\ts -0.594) & (0,\ts 0)\\
                        (0.0169,\ts -0.241) & (0.9999,\ts 0.072) & (0,\ts 0)\\
                        (0,\ts 0) & (0,\ts 0) & (1,\ts 0.072)\ea\right) 
                        \nn\\ \nn \\
D_L &=&\left(\ba{lll}   (0.978,\ts 0.094) & (0.207,\ts 0.078) & (0.00486,\ts
                          -0.261)\\ 
                        (0.207,\ts 1.094) & (0.977,\ts 0.071) & (0.0414,\ts
                        0.071)\\ 
                        (0.00745,\ts -0.093) & (0.0410,\ts 1.078) &
                        (0.999,\ts 0.072) \ea\right)
                        \nn\\ \\
U_R &=&\left(\ba{lll}   (0.976,\ts -0.241) & (0.217,\ts 0.072) & (0,\ts 0)\\
                        (0.217,\ts 0.759) & (0.976,\ts 0.072) & (0,\ts 0)\\
                        (0,\ts 0) & (0,\ts 0) & (1,\ts 0.072)\ea\right)
                        \nn\\ \nn\\ 
D_R &=&\left(\ba{lll}   (0.982,\ts 0.760) & (0.188,\ts -0.262) &(0.000237,\ts
                         -0.262)\\
                        (0.188,\ts 0.094) & (0.982,\ts 0.071) &(0.00120,\ts
                         0.065)\\  
                     (9\times 10^{-6},\ts -0.425) & (0.00122,\ts 0.745)
                     &(1,\ts -0.262) \ea\right) \ts.\nn
\eea
Finally, the CKM matrix is
\be\label{eq:CKMmodelc}
V = \left(\ba{lll}   (0.976,\ts 0) & (0.216,\ts 0) & (0.00455,\ts -0.311)\\
                        (0.216,\ts 1) & (0.976,\ts 1.1\times 10^{-5}) &
                        (0.0415,\ts 0)\\ 
                        (0.00745,\ts -0.164) & (0.0410,\ts 1.006) &
                        (0.999,\ts 0)
                        \ea\right) \ts.
\ee
This corresponds to the angles and phase
\be\label{eq:CKMangsmodelc}
\theta_{12} = 0.218 \ts, \ts\ts\ts \theta_{23} = 0.0415 \ts, \ts\ts\ts
\theta_{13} = 0.00455 \ts, \ts\ts\ts \delta_{13} = 0.977\ts.
\ee

\section*{II.9 Summary and Conclusions}

We have presented a dynamical picture of CP nonconservation arising from
vacuum alignment in extended technicolor theories. This picture leads
naturally to a mechanism involving rational phases of the alignment
matrices, evading strong CP violation without the need for a light axion or a
massless up quark. We derived complex quark mixing matrices from
ETC/TC2--based constraints on the primordial mass matrices $\CM_u$ and
$\CM_d$. These led to very realistic--looking CKM matrices. We categorized
4--quark contact interactions arising from ETC and TC2 and proposed a scheme
for estimating the strengths of these interactions. Putting this together
with the quark mixing matrices, we calculated the contributions to the
CP--violating parameters $\epsilon$, and $\sin(2\beta)$ for six sample
$\CM_q$. Only one example fit the data, but it fit quite well. Future work
will include calculating $\epsilon'/\epsilon$. It is also important to
determine the magnitude of the electroweak and top--condensate corrections to
the rational phases.

\bigskip

\nin In closing these lectures, I leave you with some thoughts of a
distinguished Harvard biologist, words that apply as well to our science:

\vskip0.10truein

\noindent {\it I have, indeed, looked into my own science's history and find a
superabundance of theorizing about anomalies. The problem is not want of a
theory but a want of evidence. If scientific advance really came from
theorizing, natural scientists would have long ago wrapped up their affairs
and gone on to more interesting matters.}

\vskip0.25truein

\hskip0.5truein --- Richard Lewontin

\hskip0.72truein    New York Review of Books

\hskip0.72truein   {\underbar {XLII}}, \#10, 69 (June 8, 1995)

\section*{Acknowledgements}

I wish to thank Genevieve Belanger for inviting me to speak at l'Ecole de
GIF. The school and the students were excellent. I enjoyed them despite my
limited French. I thank Sekhar Chivukula, Estia Eichten, and Chris Hill for
numerous influential conservations and instructions over many years. I am
grateful to Fermilab and its Theory Group for a Frontier Fellowship which
supported the writing of this research. My research was also supported by the
U.S.~Department of Energy under Grant~No.~DE--FG02--91ER40676.  L'Ecole de
GIF took place during the week September 10--14, 2001. I can never fully
express my gratitude for the kindness and generosity of spirit everyone
showed me, my wife, and countrymen that terrible week. In addition to
Genevieve and her colleagues whose names I never fully caught, I especially
want to thank Marie--Pascale Colace for her boundless help and friendship and
Eric Pilon for showing me where to bike. I would have been been lost without
those rides, especially the one up to Col de Leschaux on September~12. Carla
and I also thank Bob Cahn for his invaluable companionship during that week.

\section*{Appendix. ETC Gauge Boson Mass Scales}

To set the ETC mass scales that enter $\chetc$ in Eq.~(\ref{eq:HETC}), we
follow the rule stated in Section~II.7: The ETC scale $\METC/\getc$ in a term
involving weak eigenstates of the form $\ol q^{\ts \prime}_i q'_j \ol q^{\ts
\prime}_j q'_i$ or $\ol q^{\ts \prime}_i q'_i \ol q^{\ts \prime}_j q'_j$ (for
$q'_i = u'_i$ or $d^{\ts \prime}_i$) is approximately the same as the scale
that generates the $\ol q^{\ts \prime}_{Ri} q'_{Lj}$ mass term,
$(\CM_q)_{ij}$. To estimate quark masses in terms of $\METC/\getc$, we again
assume a model in which quarks couple via ETC to $N$ identical doublets of
technifermions transforming as $(\Ntc,1)$ under $\sutc\otimes\suc$. The
technipion decay constant (which sets the technicolor energy scale) is then
$F_T = F_\pi/\sqrt{N}$, where $F_\pi = 246\,\gev$.

The ETC gauge boson mass $\METC(q)$ giving rise to a quark mass
$m_q(\METC)$ ---an element or eigenvalue of $\CM_q$---is given by
Eq.~(\ref{eq:masses}):
\bea
\hskip1.45in 
m_q(\METC) \simeq 2{g^2_{ETC} \over {M^2_{ETC}(q)}}
\langle \ol T_L T_R \rangle_{ETC} \ts. \nn
\hskip1.19in (3)
\eea
Here, the quark mass and the technifermion bilinear condensate, $\condetc$,
are renormalized at the scale $\METC(q)$. This condensate is related to the
one renormalized at the technicolor scale $\LTC \simeq F_T$ by
Eq.~(\ref{eq:condrenorm}):
\bea
\hskip1.10in
\condetc = \condtc \ts \exp\left(\int_{\LTC}^{M_{ETC}} \ts {d \mu
\over {\mu}} \ts \gamma_m(\mu) \right) \ts, \nn
\hskip0.82in (6)
\eea
where, scaling from QCD, we expect the TC--scale condensate in Eq.(5):
\bea
\hskip1.25in
\langle \ol T_L T_R \rangle_{TC} \equiv \half \Delta_T \simeq 2 \pi F^3_T =
2\pi F^3_\pi/N^{3/2} \ts. \nn
\hskip1.07in (5)
\eea

In a walking technicolor theory the coupling $\atc(\mu)$ decreases
very slowly from its critical chiral symmetry breaking value at $\LTC$, and
$\gamma_m(\mu) \simeq 1$ for $\LTC \simle \mu \simle \METC$.
An accurate evaluation of the condensate enhancement integral in
Eq.~(\ref{eq:condrenorm}) requires detailed specification of the technicolor
model and knowledge of the $\beta(\atc)$--function for large
coupling.\footnote{See Ref.~\cite{multiklrm} for an attempt to calculate this
integral in a walking technicolor model.} Lacking this, we estimate the
enhancement by assuming that
\bea\label{eq:gmmb}
\gamma_m(\mu) &= \left\{\ba{ll} 1 & {\rm for} \ts\ts\ts\ts \LTC < \mu <
  \CM_{ETC}/\kappa^2 \\
0 & {\rm for} \ts\ts\ts\ts \mu > \CM_{ETC}/\kappa^2
\ea \right.
\eea
Here, $\CM_{ETC}$ is the largest ETC scale, i.e., the one generating the
smallest term in the quark mass matrix for $\kappa =1$. The number $\kappa >
1$ parameterizes the departure from the strict walking limit (i.e., $\gamma_m
= 1$ constant all the way up to $\CM_{ETC}/\kappa^2$). Then, using
Eqs.~(\ref{eq:masses}) and~(\ref{eq:condrenorm}), we obtain
\bea\label{eq:metc}
{\METC(q)\over{\getc}} &= \left\{\ba{ll} {\sqrt{64\pi^3\alpha_{ETC}}\ts
    F^2_\pi\over {N m_q}}  & {\rm if} \ts\ts\ts\ts \METC(q) <
  \CM_{ETC}/\kappa^2 \\ \\
  \sqrt{{4\pi \CM_{ETC} F^2_\pi \over{\kappa^2 N m_q}}} & {\rm if}
  \ts\ts\ts\ts
  \METC(q) > \CM_{ETC}/\kappa^2
\ea \right.
\eea
To evaluate this, we take $\alpha_{ETC} = 3/4$, a moderately strong value as
would be expected in walking technicolor~\cite{multiklrm}, and $N = 10$, a
typical number of doublets in TC2 models with topcolor
breaking~\cite{tctwokl}. Then, taking the smallest quark mass at the ETC
scale to be $10\,\mev$, we find $\CM_{ETC} = 7.17\times 10^4\,\tev$. The
resulting estimates of $\METC/\getc$ were plotted in Figure~2 for $\kappa =
1, \sqrt{10}$, and 10. They run from $\METC/\getc = 46\,\tev$ for $m_q =
5\,\gev$ to $2.33/\kappa\times 10^4\,\tev/$ for $m_q = 10\,\mev$. Very
similar results are obtained for $\alpha_{ETC} = 1/2$ and $N=8$.

\vfil\eject

\vfil\eject

\end{document}